\documentclass[fleqn,usenatbib,useAMS]{mnras}
\usepackage{aas_macros}
\usepackage{bibentry}
\usepackage[T1]{fontenc}
\usepackage{amsmath}
\usepackage{mathtools}
\usepackage{float}
\usepackage{amssymb}
\usepackage{epsfig}
\usepackage{color}
\usepackage{ulem}
\usepackage{subfig}
\usepackage{graphicx}
\usepackage{bm}
\usepackage{syntonly}

\newcommand{\gsim}{\lower.7ex\hbox{$\;\stackrel{\textstyle>}{\sim}\;$}}
\newcommand{\lsim}{\lower.7ex\hbox{$\;\stackrel{\textstyle<}{\sim}\;$}}

\title[Galaxy Shape]{Evolution of Galaxy Shapes from Prolate to Oblate through Compaction Events}
\author[Tomassetti et al.]
{\parbox[t]{\textwidth} 
{
Matteo Tomassetti$^{1,2}$\thanks{E-mail: matteo.tomassetti@mail.huji.ac.il},
Avishai Dekel$^{1}$\thanks{E-mail: dekel@huji.ac.il},
Nir Mandelker$^{1}$,
Daniel Ceverino$^{3}$, Sharon Lapiner$^{1}$, Sandra Faber$^{2}$, Omer Kneller$^{1}$, Joel Primack$^{2}$, Tanmayi Sai$^{2}$
\\
}
\\
\\
$^1$Center for Astrophysics and Planetary Science,
Racah Institute of Physics, The Hebrew University, Jerusalem 91904, Israel\\
$^2$ Department of Physics, University of California, Santa Cruz, CA 95064,
USA\\
$^3$ Universit{\"a}t Heidelberg, Zentrum f{\"u}r Astronomie, Institut f{\"u}r Theoretische Astrophysik, Albert-Ueberle-Str. 2, D-69120 Heidelberg, Germany
}

\begin{document}

\date{Accepted 2016 March 9. Received 2016 March 9; in original form 2015 December 19}

\large

\pagerange{\pageref{firstpage}--\pageref{lastpage}} \pubyear{2002}

\maketitle

\label{firstpage}

\begin{abstract}
We study the evolution of global shapes of galaxies using cosmological simulations.
The shapes refer to the components of dark matter (DM), stars and gas at the stellar half-mass radius. Most galaxies undergo a characteristic compaction event into a blue nugget at $z\sim2-4$, which marks the transition from a DM-dominated central body to a self-gravitating baryonic core.
We find that in the high-$z$, DM-dominated phase, the stellar and DM systems tend to be triaxial, preferentially prolate and mutually aligned. The elongation is supported by an anisotropic velocity dispersion that originates from the assembly of the galaxy along a dominant large-scale filament.
We estimate that torques by the dominant halo are capable of inducing the elongation of the stellar system and its alignment with the halo. Then, in association with the transition to self-gravity, small-pericenter orbits puff up and the DM and stellar systems evolve into a more spherical and oblate configuration, aligned with the gas disc and associated with rotation.
This transition typically occurs when the stellar mass is $\sim 10^9$ M$_\odot$ and the escape velocity in the core is $\sim 100$ km s$^{-1}$, indicating that supernova feedback may be effective in keeping the core DM-dominated and the system prolate. The early elongated phase itself may be responsible for the compaction event, and the transition to the oblate phase may be associated with the subsequent quenching in the core.
\end{abstract}


\section{Introduction}
The global shape of a galaxy is a major characteristic of the galaxy, which reflects the dominant mechanisms involved in its formation and evolution. In particular, at low redshifts, galaxies tend to be oblate systems, either discs or puffed-up oblate spheroids. These shapes reflect a basic difference in the kinematics, from rotation-dominated systems to pressure-supported spheroids, with a varying ratio of rotation velocity to velocity dispersion.

\smallskip
There is a very surprising indication from observations that galaxies of relatively low masses at high redshift are actually elongated, preferentially prolate systems. 
Based on the observed ellipticities of projected images, \citet{Law+12} found that the population of galaxies at $z=1.5-3.6$ is better described by triaxial systems, with a ratio between the minor and major semi-axes of $c/a\sim0.3$ and a ratio between the intermediate and major semi-axis of $b/a\sim0.7$. \citet{Chang+13} looked at a
sample of early-type galaxies at $1<z<2.5$, and adopting ad hoc  assumptions concerning the distributions of the three-dimensional axial ratios,  concluded that the oblate fraction in the three-dimensional shapes is increasing with cosmic time for galaxies with $M_*<10^{10.5}$ M$_\odot$.
\citet{vanderWel+14} deduced in a similar way from CANDELS+3D-$HST$ data that most of the galaxies with stellar mass $M_*\sim10^{8.5}-10^{9.5}$ M$_\odot$ at $z\sim1-2$ are prolate, while more massive galaxies, or galaxies at lower redshift, could be oblate. \citet{Takeuchi+15} confirmed this finding using GOOD-S and SXDS data. They deduced intrinsic triaxial shapes where $b/a$ evolves from  $\sim0.81\pm0.04$ at $z\sim2$ to $b/a\sim0.92\pm0.05$ at $z\sim0.7$. This is consistent with an evolution from a more prolate configuration to a rounder, more oblate system. Our goal in this paper is to measure the intrinsic shape evolution in cosmological simulations, quantify its dependence on mass and time, and attempt to understand the origin of prolate galaxies at high redshift and their evolution into oblate systems in more massive galaxies at low redshift.

\smallskip
Based on $N$-body simulations of structure formation in the standard $\Lambda$ cold dark matter ($\Lambda CDM$) cosmology, we know that the inner parts of dark-matter (DM) haloes soon after their assembly show triaxial shape, with average axial ratios at the virial radius of $b/a\sim0.6$ and $c/b\sim0.67$ (where the ellipsoidal semi-axes are $a\ge b\ge c$), namely tending towards prolate shapes  \citep[e.g.][]{Frenk+88,Dubinski+91,Jing+02,Bailin+05,Allgood+06} .

\smallskip
The origin of this prolateness is likely to be the assembly of the halo along a preferred direction, defined by the dominant large-scale filament within which the halo is formed by mergers of smaller haloes  and smoother streaming. This induces an anisotropic velocity dispersion  within the halo, which support a prolate shape with the major axis  aligned with the large-scale filament \citep{Codis+15,Laigle+15}. The suggested correlation between the shapes on galactic scales and on the scale of the extended cosmic-web environment should be addressed.

\smallskip
The shape of the stellar system may be similar and aligned with the DM-halo shape for the same reason, especially for the stars that entered the galaxies with the same mergers that built the DM halo and along the same filament. Moreover, especially for younger stars that form within the elongated halo, if the inner halo dominates the potential well, torques exerted by the halo may induce a similar elongated shape in the stellar system, and make it align with the halo elongation. Whether these torques are indeed capable of inducing the shape of the stellar system is an open question that should be addressed.

\smallskip
A central concentration of baryons that dominates the potential at the halo centre may in turn affect the DM-halo and stellar shapes, and make them rounder  \citep[e.g., following the analysis of][]{Athanassoula+05,Debattista+08} . The box orbits that characterize the triaxial configuration can be deflected and puffed up by the central mass, thus making the system oblate and rounder.

\smallskip
Qualitatively, the central mass has to be $\sim20$ per cent of the disk mass to destroy a bar
\citep{Shen+04,Debattista+06}. This indicates that in the region where the central baryonic mass dominates the potential it may be capable of affecting the inner DM-halo and stellar shapes. This should be demonstrated using simulations.

\smallskip
The evolution of shape may be tightly linked to the characteristic sequence of events that galaxies undergo, typically in the redshift range $z=4-2$, based on both theory and observations. Triggered by mergers, counter-rotating streams, violent disc instability or other mechanisms, the galaxy goes through a wet compaction process and forms a compact, star-forming ``blue nugget". This leads to an inside-out quenching process into a compact, passive ``red nugget", the likely progenitor of the centre of an early-type galaxy today.
The observational basis for this picture is becoming solid, both for the red-nugget phenomenon  \citep{Trujillo+06a,Trujillo+06b,vanDokkum+08,Damjanov+09,Damjanov+11,Newman+10,Bruce+12,Whitaker+12,vanDokkum+14} and for their potential blue-nugget progenitors \citep{Barro+13,Barro+14a,Barro+14b,Bruce+14,Nelson+14,Williams+14,Tacchella+15}. The theory, partly based on the same simulations that we analyse here, is also becoming robust \citep{Dekel+14,Tacchella+15b, Tacchella+15c,Zolotov+15}.

\begin{figure}
\begin{center}
\includegraphics[scale=0.4]{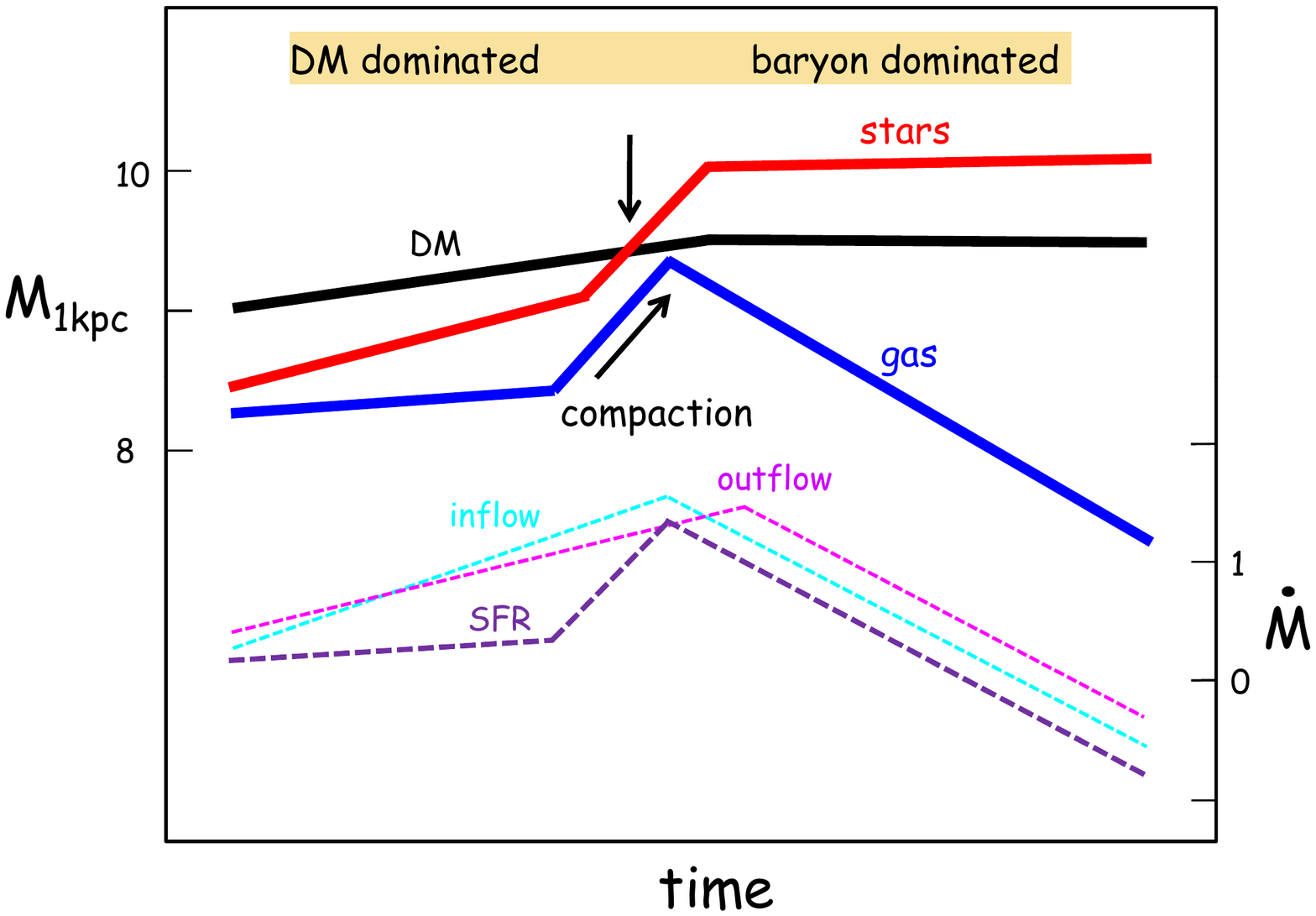}
\caption{\label{fig:fig1_zolotov} Characteristic evolution of masses and their rates of change within the central regions of the galaxies in our simulations [according to the analysis of \citet{Zolotov+15} and \citet{Tacchella+15b}]. After an early phase of gradual mass growth and star formation, there is a well-defined, relatively short phase of wet compaction in the inner 1 kpc, reaching a peak of central gas density and SFR (a blue nugget). After that, there is a longer phase of gas depletion and quenching of SFR caused by a low rate of inflow to the centre compared to the sum of SFR and outflow rate. The result is a compact quenched galaxy (a red nugget), where the central stellar density remains roughly constant from the blue-nugget phase and one. The $y$-axes are log-scaled, and the mass and its rate of change are in $M_\odot$ and $M_\odot$ yr$^{-1}$, respectively. The ticks on the $y$-axes are mass- and redshift dependent, and the plot refers to a typical case where the compaction is at $z\sim2-3$ and the post-compaction stellar mass within 1 kpc is $\sim~10^{10}$ M$_\odot$.}
\end{center}
\end{figure}

\smallskip
A very relevant finding from the simulations \citep[][Fig. 2-4]{Zolotov+15} is that the major compaction event is associated with a transition from a situation where the mass in the central region of the galaxy (either the inner half-mass, or inside 1 kpc) is dominated by the DM to a configuration where the central mass is dominated by the compact nugget of baryons. Fig. \ref{fig:fig1_zolotov}, following \citet{Zolotov+15}, shows a cartoon that illustrates the typical evolution through compaction and the associated transition of dominance at the center.

\smallskip
As alluded to in the above discussion, the transition from central DM dominance to baryon dominance may induce a transition of shape from prolate to oblate systems. Indeed, \citet{Ceverino+15}, focusing on five simulated galaxies of relatively low mass of $\sim 10^9$ M$_\odot$ at $z\sim2$, identified a tendency for a prolate shape for the halo and the stellar system. These five simulations were part of the large suite used here. The prolate systems are indeed found in snapshots of these galaxies where the core \footnote{When referring to the core of a galaxy we mean the innermost 1 kpc.} is dominated by DM, typically before the compaction phase of evolution, which for these galaxies typically occurs after $z\sim2$.

\smallskip
In this paper we use a suite of 34 galaxies simulated at high resolution in a cosmological setting, as described in the following section, to address the shape evolution in a systematic way.
Our first goal is to study the evolution of shape for the components of DM, stars and gas. We will then attempt to verify the relation between the shape and the DM-to-baryon mass ratio in the central regions, and study the correlation between the evolution of shape and the evolution of the galaxy though the compaction phase. We will address in particular the characteristic conditions for the transition in terms of mass and redshift. Once these correlations are established in the simulations, we will make first attempts at understanding the mechanisms responsible for the stellar-system elongation in terms of the large-scale filamentary structure versus torques exerted by the halo, and the way the baryonic central concentration eventually leads to oblate systems.

\smallskip
This paper is organized as follows. In Section \ref{sec:simulations}, we describe the simulation suite used in this work. In Section \ref{sec:measuring_shape}, we describe the method used to measure the global shape of a three-dimensional system, and compare to other methods. In Section \ref{sec:shape_in_pictures}, we illustrate using projected images the application of the shape measure in a few cases, and the typical evolution of shapes for these galaxies. In Section \ref{sec:evolution_of_shape}, we study the evolution of shape for the whole simulation suite and address the correlation with the evolution of the DM-to-baryon mass ratio and the major compaction event. In Section \ref{sec:alignment}, we discuss the relative alignment among the different components, between physical shape and the anisotropy of the velocity dispersion, and between galactic scales and the large-scale of the cosmic web. In Section \ref{sec:theoretical_interpretation},  we discuss issues concerning the origin of elongation at high redshift and the change of shape at lower redshift. Finally, in Section \ref{sec:conclusions}, we summarize our conclusions and discuss them.

\section{Simulations}\label{sec:simulations}
\subsection{Zoom-in cosmological simulations}

\smallskip
We use the VELA suite of simulations with radiation pressure (generation 3), which consists of 34 zoom-in cosmological simulations with maximum physical spatial resolution ranging between 17.5 and 35 pc. Many of these galaxies have been evolved to a final redshift of $z_{\rm fin}=1$ and they encompass a range of masses. For instance, at $z\sim2$, the virial masses of the DM haloes range from $10^{11}$ to $10^{12}$ M$_\odot$ and the stellar masses of the galaxies are between $10^9$ and  $6\times10^{10}$ M$_\odot$.

\smallskip
The simulations have been performed with the Adaptive Refinement Tree (ART) code \citep{Kravtsov+97,Ceverino+09}, which solves the hydrodynamics equations and gravity on a Eulerian adaptive mesh. The code also includes gas cooling by atomic hydrogen and helium, metal and molecular hydrogen, and heating by the ultraviolet (UV) background with partial self-shielding, star formation, stellar mass loss, metal enrichment of the interstellar medium (ISM) and stellar feedback.

\smallskip
Supernovae and stellar winds are implemented by local injection of thermal energy as in \citet{Ceverino+09,Ceverino+10,Ceverino+12}. Radiative stellar feedback takes into account the effect of radiation pressure from ionizing photons, and it is implemented following the theoretical estimates of \citet{Dekel+13b}, as described in \citet{Ceverino+14}.

\smallskip
In the following, we give a few more details for the relevant subgrid physics but for a more complete description we refer to \citet{Ceverino+14}.
Cooling and heating rates are tabulated for a given
gas density, temperature, metallicity and UV background
based on the CLOUDY code \citep{Ferland+98}, assuming a slab of thickness 1 kpc. A uniform UV background based on the redshift-dependent \citet{Haardt+96} mode is assumed, except at gas densities higher than 0.1 cm$^{-3}$, where a substantially suppressed UV background is used ($5.9\times10^6$ erg s$^{-1}$ cm$^{-2}$ Hz$^{-1}$) in order to mimic the partial self-shielding of dense gas, allowing dense gas to cool down to temperatures of 300 K. The assumed equation of state is that of an ideal mono-atomic gas. Artificial fragmentation on the cell size is prevented by introducing a pressure floor,
which ensures that the Jeans scale is resolved by at least seven
cells \citep[see][]{Ceverino+10}.

\smallskip
Star formation is assumed to occur at densities above a
threshold of 1 cm$^{-3}$ and at temperatures below $10^4$ K. Most
stars ($> 90$ per cent) form at temperatures well below $10^3$ K, and
more than half of the stars form at 300 K in cells where the
gas density is higher than 10 cm$^{-3}$. The code implements a
stochastic star formation model that yields a star formation
efficiency per free-fall time of 2 per cent, using the stellar initial
mass function of \citet{Chabrier+03}. At the given resolution,
this efficiency roughly mimics the empirical Kennicutt-
Schmidt law \citep{Kennicutt+98}.

\smallskip
The thermal stellar feedback model incorporated in the
code releases energy from stellar winds and supernova explosions
at a constant heating rate over 40 Myr following
star formation. The heating rate due to feedback may or
may not overcome the cooling rate, depending on the gas
conditions in the star-forming regions \citep{Dekel+86,Ceverino+09}. There is no artificial shutdown
of cooling implemented in these simulations. Runaway stars
are included by applying a velocity kick of about 10 km s$^{-1}$ to
30 per cent of the newly formed stellar particles. The code also
includes the later effects of Type Ia supernova and stellar
mass loss, and it follows the metal enrichment of the
ISM.

\smallskip
We incorporate radiation pressure through the addition
of a non-thermal pressure term to the total gas pressure
in regions where ionizing photons from massive stars
are produced and may be trapped. This ionizing radiation
injects momentum in the cells neighbouring massive star
particles younger than 5 Myr, and whose column density exceeds
10$^{21}$ cm$^{-2}$, isotropically pressurizing the star-forming
regions (as described also in \citet{Agertz+13}).

\smallskip
The initial conditions for the simulations are based on
DM haloes that were drawn from dissipationless $N$-body
simulations at lower resolution in three large comoving
cosmological boxes.  We assume the standard $\Lambda$CDM cosmological
model with the WMAP5 values of the cosmological
parameters, namely $\Omega_m=0.27$, $\Omega_\Lambda=0.73$, $\Omega_b=0.045$, $h=0.7$ and $\sigma_8=0.82$ \citep{Komatsu+09}. We selected
each halo to have a given virial mass at $z=1$ and no ongoing
major merger at $z=1$. This latter criterion eliminates less
than 10 per cent of the haloes which tend to be in a dense environment
at $z\sim1$, and it induces only a minor selection effect
at higher redshifts. The virial masses at $z=1$ were chosen to be in the range $M_{\rm vir}=2\times10^{11}-2\times10^{12}$ M$_\odot$, about a median of $4.6\times10^{11}$ M$_\odot$. If left in isolation, the median mass
at $z=0$ was intended to be $10^{12}$ M$_\odot$. Realistically, the
actual mass range is broader, with some of the haloes merging
into more massive haloes that eventually host groups at
$z=0$.

\subsection{Global properties of the galaxy sample}

In Table \ref{tab:sim_table}, we list the virial masses, the virial radii, the galaxy stellar masses and the effective radii of the entire sample of 34 galaxies. The virial mass $M_{\rm vir}$ is the total mass within a sphere of radius $R_{\rm vir}$, which encompasses an overdensity of $\Delta(z)=(18\pi^2-82\Omega_\Lambda(z)-39\Omega_\lambda(z)^2)/\Omega_m(z)$, where $\Omega_\Lambda(z)$ and $\Omega_m(z)$ are the cosmological parameters at $z$ \citep{Bryan+98,Dekel+06}. The stellar mass $M_*$ is measured within a sphere of radius 10 kpc about the galaxy center. The effective radius $R_{\rm e}$ is the three-dimensional half-mass radius corresponding to this $M_*$.

\smallskip
Every galaxy is analyzed at outputs separated by a constant interval in the expansion factor $\Delta a=0.01$ from $a\sim0.125$ ($z=7$) until $a_{\rm fin}$ ($z_{\rm fin}$). The total sample consists of $\sim1100$ snapshots.

\begin{table*}
\centering
    \begin{tabular}{ccccccccccccc}
\hline
\hline
Galaxy & $M_{\rm vir}$ & $M_*$ & $R_{\rm vir}$ & $R_{\rm e}$
   & $M_{\rm vir}$ & $M_*$ & $R_{\rm vir}$ & $R_{\rm e}$&$a_{\rm fin}$&$z_{\rm fin}$ \\
       & $10^{12}M_\odot$ & $10^{10}M_\odot$  & kpc & kpc
   & $10^{12}M_\odot$ & $10^{10} M_\odot$  & kpc  & kpc &\\
       & $(z=2)$ & $(z=2)$ &$(z=2)$ & $(z=2)$&
    $(z_{\rm fin})$ & $(z_{\rm fin})$ & $(z_{\rm fin}$) &$(z_{\rm fin}$) &&\\
\hline
V01 & 0.16 & 0.22 & 58.25 & 1.06 & 0.48 & 1.51 & 123.75 & 2.18 & 0.50 & 1.00 \\
V02 & 0.13 & 0.19 & 54.50 & 2.19 & 0.39 & 0.92 & 115.25 & 2.09 & 0.50 & 1.00 \\
V03 & 0.14 & 0.43 & 55.50 & 1.70 & 0.32 & 1.00 & 108.00 & 1.91 & 0.50 & 1.00 \\
V04 & 0.12 & 0.10 & 53.50 & 2.15 & 0.18 & 0.36 & 89.75 & 1.14 & 0.50 & 1.00 \\
V05 & 0.07 & 0.10 & 44.50 & 2.66 & 0.15 & 0.28 & 83.00 & 2.46 & 0.50 & 1.00 \\
V06 & 0.55 & 2.16 & 88.25 & 1.06 & 0.75 & 2.57 & 108.75 & 1.13 & 0.37 & 1.70 \\
V07 & 0.90 & 5.67 & 104.25 & 2.78 & 1.53 & 7.07 & 196.50 & 3.37 & 0.50 & 1.00 \\
V08 & 0.28 & 0.35 & 70.50 & 0.76 & 1.37 & 3.37 & 196.50 & 3.40 & 0.50 & 1.00 \\
V09 & 0.27 & 1.06 & 70.50 & 1.82 & 0.21 & 4.18 & 103.00 & 1.47 & 0.40 & 1.50 \\
V10 & 0.13 & 0.64 & 55.25 & 0.54 & 0.75 & 2.38 & 158.50 & 0.79 & 0.50 & 1.00 \\
V11 & 0.27 & 0.91 & 69.50 & 2.99 & 0.44 & 1.55 & 132.50 & 3.12 & 0.46 & 1.17 \\
V12 & 0.27 & 2.03 & 69.50 & 1.22 & 0.25 & 2.22 & 110.50 & 1.32 & 0.44 & 1.27 \\
V13 & 0.31 & 0.69 & 72.50 & 3.21 & 0.87 & 2.05 & 158.25 & 4.25 & 0.40 & 1.50 \\
V14 & 0.36 & 1.30 & 76.50 & 0.35 & 0.21 & 2.78 & 104.25 & 0.70 & 0.41 & 1.44 \\
V15 & 0.12 & 0.56 & 53.25 & 1.31 & 0.35 & 1.04 & 123.25 & 1.94 & 0.50 & 1.00 \\
V16 & - & - & - & - & 0.50 & 4.29 & 62.75 & 0.69 & 0.24 & 3.17 \\
V17 & - & - & - & - & 1.13 & 8.41 & 105.75 & 1.33 & 0.31 & 2.23 \\
V19 & - & - & - & - & 0.88 & 4.50 & 91.25 & 1.23 & 0.29 & 2.45 \\
V20 & 0.53 & 3.70 & 87.50 & 1.81 & 1.06 & 6.86 & 146.25 & 3.74 & 0.44 & 1.27 \\
V21 & 0.62 & 4.10 & 92.25 & 1.76 & 0.86 & 5.74 & 151.50 & 3.53 & 0.50 & 1.00 \\
V22 & 0.49 & 4.45 & 85.50 & 1.32 & 0.62 & 4.51 & 136.00 & 1.92 & 0.50 & 1.00 \\
V23 & 0.15 & 0.83 & 57.00 & 1.38 & 0.47 & 2.51 & 123.00 & 1.98 & 0.50 & 1.00 \\
V24 & 0.28 & 0.92 & 70.25 & 1.79 & 0.36 & 2.15 & 108.25 & 1.73 & 0.48 & 1.08 \\
V25 & 0.22 & 0.73 & 65.00 & 0.82 & 0.32 & 1.39 & 108.00 & 1.11 & 0.50 & 1.00 \\
V26 & 0.36 & 1.60 & 76.75 & 0.77 & 0.42 & 2.14 & 120.00 & 1.97 & 0.50 & 1.00 \\
V27 & 0.33 & 0.80 & 75.50 & 2.45 & 0.35 & 1.86 & 114.50 & 4.00 & 0.50 & 1.00 \\
V28 & 0.20 & 0.24 & 63.50 & 3.23 & 0.22 & 0.50 & 96.00 & 1.64 & 0.50 & 1.00 \\
V29 & 0.52 & 2.34 & 89.25 & 1.96 & 0.90 & 3.34 & 152.50 & 2.78 & 0.50 & 1.00 \\
V30 & 0.31 & 1.66 & 73.25 & 1.56 & 0.32 & 1.67 & 76.25 & 1.64 & 0.34 & 1.94 \\
V31 & - & - & - & - & 0.23 & 0.85 & 38.50 & 0.51 & 0.19 & 4.26 \\
V32 & 0.59 & 2.68 & 90.50 & 2.60 & 0.59 & 2.68 & 90.50 & 2.60 & 0.33 & 2.03 \\
V33 & 0.83 & 4.80 & 101.25 & 1.22 & 1.46 & 8.92 & 143.75 & 1.64 & 0.39 & 1.56 \\
V34 & 0.52 & 1.61 & 86.50 & 1.90 & 0.62 & 1.90 & 97.00 & 2.06 & 0.35 & 1.86 \\
V35 & - & - & - & - & 0.13 & 0.56 & 37.50 & 0.67 & 0.22 & 3.55 \\
\hline
\hline
\end{tabular}
\caption{The suite of 34 simulated galaxies.
The galaxy name Vxx is short for VELA\_V2\_xx.
Quoted are the
total mass $M_{\rm vir}$, the stellar mass $M_*$, the virial radius $R_{\rm vir}$ and the
effective stellar (half-mass) radius $R_{\rm e}$
both at $z=2$ and at the final simulation snapshot,
$a_{\rm fin}=(1+z_{\rm fin})^{-1}$.
}
\label{tab:sim_table}
\end{table*}

\subsection{Limitations of the simulations}

These simulations are state-of-the-art in terms of the high-resolution
AMR hydrodynamics and the treatment of key physical processes at the
subgrid level. In particular, they properly trace the cosmological streams
that feed galaxies at high redshift, including mergers and smooth flows,
and they resolve the violent disc instability that governs the high-$z$ disc
evolution and the bulge formation \citep{Ceverino+10,Ceverino+12,Ceverino+15,Mandelker+14}.

\smallskip
Like in other simulations,
the treatment of star formation and feedback processes can still be improved.
Currently,
the code assumes a star formation rate (SFR) efficiency per free-fall time that is more
realistic than in earlier versions,
it does not yet follow in great detail the formation of molecules
, subtle the effect of metallicity on SFR \citep{Krumholz+12}, and the detailed suppression of SFR by photoelectric effect on dust grains \citep{Forbes+15}.
Furthermore, the resolution does not allow the capture of the Sedov-Taylor
adiabatic phase of supernova feedback. 
The radiative stellar feedback assumed no infrared trapping,
in the spirit of the low trapping advocated by
\citet{Dekel+13b} based on \citet{Krumholz+12b}. 
On the other hand, other works assume more significant trapping 
\citep{Krumholz+10,Murray+10,Hopkins+12}, which makes the assumed strength of the radiative stellar feedback moderate.
Finally, AGN feedback, and feedback associated with cosmic rays and magnetic
fields, is not yet incorporated.
Nevertheless, as shown in \citet{Ceverino+14},
the SFRs, gas fractions,
and stellar to halo mass fractions are all in the ballpark of the estimates 
deduced from observations, providing a better match to observations than 
earlier versions of the ART simulations.

\smallskip
The uncertainties and any possible remaining mismatches by a factor of order 2 are comparable to the observational uncertainties. For example, our simulations produce stellar-to-halo mass ratios that are in the ballpark of the values estimated from observations (see \citet{Tacchella+15c}), and within the observational uncertainties (see the appendix in \citet{Tacchella+15c})), which are also comparable to the uncertainties associated with the feedback recipes in the simulations.

\smallskip
It seems that in the current simulations, the compaction and the subsequent onset of quenching occur at cosmological times that are consistent with observations [see fig. 12 of \citet{Zolotov+15} and fig. 2 of \citet{Barro+13}]. However, with some of the feedback mechanisms not yet incorporated (e.g., fully resolved supernova feedback and AGN feedback), full quenching to very low sSFR values may not be entirely reached in many galaxies by the end of the simulations at $z\sim1$. One should be aware of this limitation of the current simulations. Nevertheless, for all the above reasons, we adopt in this work the hypothesis that the simulations grasp the qualitative features of the main physical mechanisms that govern galaxy evolution through the processes of compaction and subsequent quenching  and the associated evolution of global shape.

\section{Measuring Shape}\label{sec:measuring_shape}

\subsection{The Shape Tensor}

A standard way to characterize the shape of a system is through its shape tensor \citep{Allgood+06}, $\mathcal{S}$, which is a symmetric matrix defined as
\begin{equation}\label{eq_def_S}
\mathcal{S}_{i,j} = \frac{1}{M}\,\sum_k\,m_k\,(\mathbf{r}_k)_i\,(\mathbf{r}_k)_j\,\,\,,
\end{equation}
where $m_k$ is the mass of the $k$-th particle (or cell), $(\mathbf{r}_k)_i$ its distance from the centre along the axis $i$ and $M$ the total mass. The eigenvalues of $\mathcal{S}$ are proportional to the squares of the semi-axes ($a>b>c$) of the ellipsoid that describes the spatial distribution of the particles (or cells) that constitute the system, and corresponding eigenvectors mark the orientations of these principal axes.

\smallskip
In particular, for a uniform-density ellipsoid, one has
\begin{equation}
\mathcal{S}=\frac{1}{\alpha}\left[\begin{array}{ccc}
a^2&0 &0 \\
0 & b^2 &0 \\
0 & 0 & c^2\\
\end{array}\right]\,\,\,,
\end{equation}
with $\alpha = 5$, while for an infinitely thin ellipsoidal shell one has $\alpha = 3$.

\smallskip
In order to compute the shape of the DM, stellar and gaseous distribution inside a spherical region of size $R$ we apply an iterative algorithm. Starting with the spherical ellipsoid $a=b=c=R$, we proceed iteratively as follows:

\begin{itemize}
\item compute $\mathcal{S}$ for all the particles within the ellipsoid (or ellipsoidal shell) and derive its eigenvalues and eigenvectors;
\item rescale the eigenvalue such that $a=R$, and rotate the system to the reference frame of the eigenvectors;
\item repeat the previous steps until convergence.
\end{itemize}

\smallskip
When the shape is determined for a thin ellipsoidal shell, we define the shell to be of a varying thickness by including the points $(x,y,z)$ that obey
\begin{equation}
1-\Delta \leq \left( \frac{x}{a} \right)^2 + \left( \frac{y}{b} \right)^2  +\left( \frac{z}{c} \right)^2  \leq 1 \, .
\end{equation}
We adopt here for the galaxy $R=R_{\rm e}$ and $\Delta=0.03$.

\smallskip
In some rare cases of a very disturbed system, the process may not converge. This may occur for the gas component at very high redshifts, when the system undergoes mergers. In such cases, we allow a maximum of 100 iterations and select the ellipsoid that has the smallest relative error among the eigenvalues of two consecutive steps.

\subsection{The Shape Parameters}

\begin{figure*}
\includegraphics[page=1,scale=0.4]{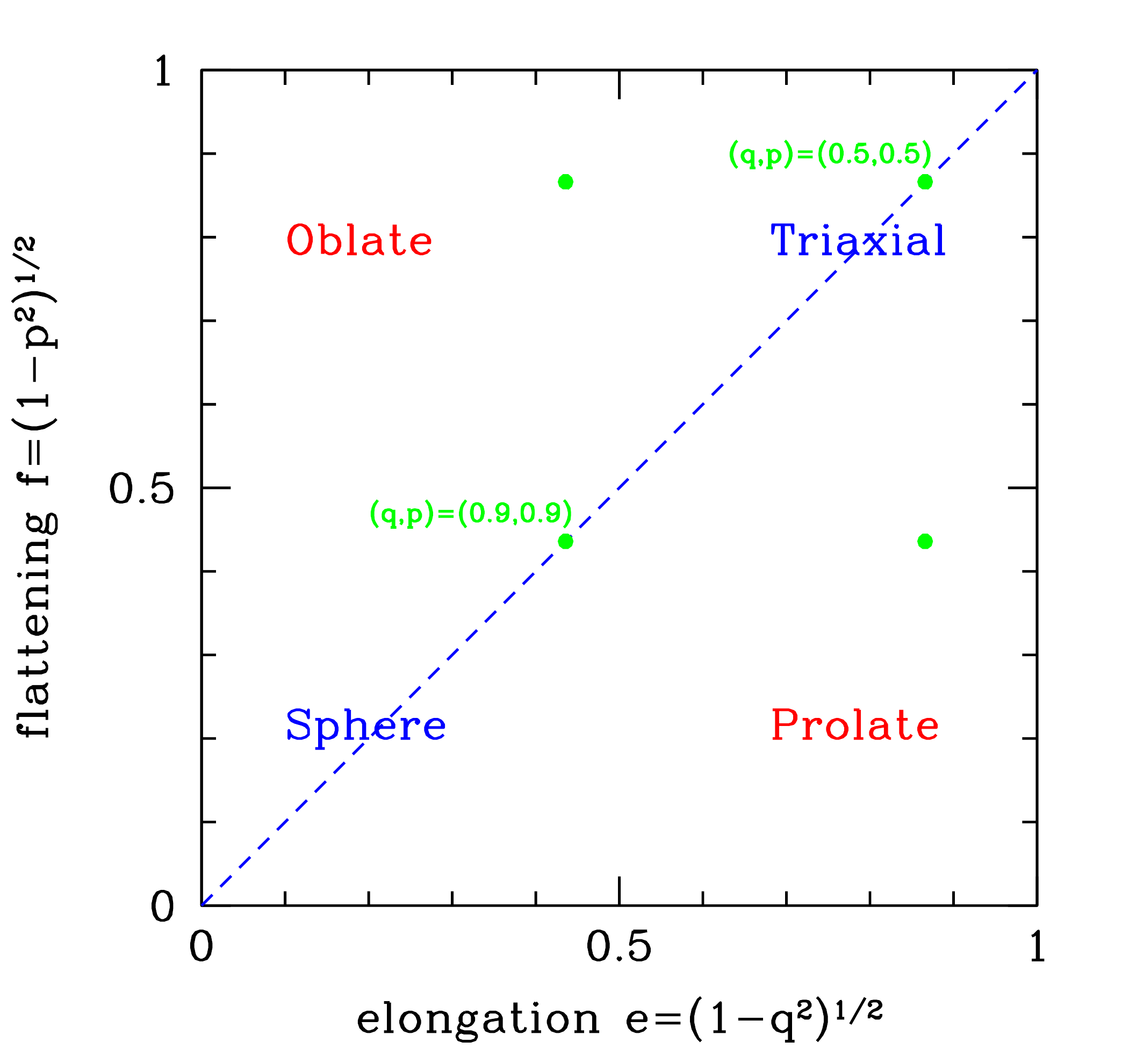}\includegraphics[page=2,scale=0.4]{shape_planes.pdf}
\caption{\label{fig:ef_plane} Presenting the shape. The left-hand panel shows the elongation versus flattening plane. The blue dashed line is the locus of pure triaxial systems, ranging from a sphere (at the bottom-left corner) to a very elongated triaxial system (at the top-right corner). The line separates  oblate and prolate ellipsoids, which lie in the top left and bottom right of the plane, respectively. The green points mark the location on the plane for four different combinations of $q$ and $p$ as indicated in parentheses. The right-hand panel shows the triaxiality versus elongation squared plane. The bottom-right half of the plane is forbidden. A sphere can obtain any value of $T$. Triaxial systems lie on a curve that is approximated by the dashed-blue line.  We adopt hereafter the presentation in the $e-f$ plane.}
\end{figure*}

The shape of an ellipsoid with axes $a\ge b\ge c$ can be characterized by two axial ratios, e.g.
\begin{align}
q = \frac{b}{a}, \qquad p =\frac{c}{b}\,\,\,,
\end{align}
each ranging from 0 to 1.

\smallskip
An ellipsoid is oblate if $q<p$ and it is prolate if $p<q$. Extreme examples are a disk where $p\ll q\sim 1$ and a filament where $q \ll p \sim 1$. When $p\sim q$ the system is triaxial; it is close to a sphere when $p \sim q \sim 1$ and it is elongated when $p\sim q \ll 1$. For the purpose of a more uniform coverage of the parameter plane, we replace $q$ and $p$ by the parameters of \textit{elongation} and \textit{flattening},
\begin{align}
e = (1-q^2)^{1/2}, \qquad  f = (1-p^2)^{1/2}\,\,\,,
\end{align}
also ranging from 1 to 0, in the opposite sense.

\smallskip
In the plane of $e$ and $f$ (as in $q$ and $p$), the different shapes occupy distinct loci. With $e$ along the $x$-axis and $f$ along the $y$-axis, oblate systems are at the top left, prolate systems are at the bottom-right, and triaxial systems are along the diagonal, ranging from spheres at the bottom left to elongated triaxial at the top right.

\smallskip
Another common choice is to replace $p=c/b$ by $s=c/a$, and use the \textit{triaxiality} parameter \citep{Franx+91}
\begin{equation}
T = \frac{1-q^2}{1-s^2}\,\,\,,
\end{equation}
also ranging from 0 to 1.
It is common to consider a system with $T<1/3$ as oblate, $1/3<T<2/3$ as triaxial, and $T>2/3$ as prolate. However, $t$ cannot fully characterize the shape by itself. For example, a near-spherical system can assume any value of $T$ between 0 and 1. One can use for example the plane $T-e$, but we note that half of this is plane is excluded as $T\ge e$.

\smallskip
The classification schemes for the shape of an ellipsoid are summarized in Fig. \ref{fig:ef_plane}. Elongation and flattening describe the shape more accurately than the elongation squared and the triaxiality since the different classes of ellipsoid occupy very distinct regions in the $e-f$ plane.

\subsection{Accuracy of our method}

\begin{figure*}
\includegraphics[scale=0.4]{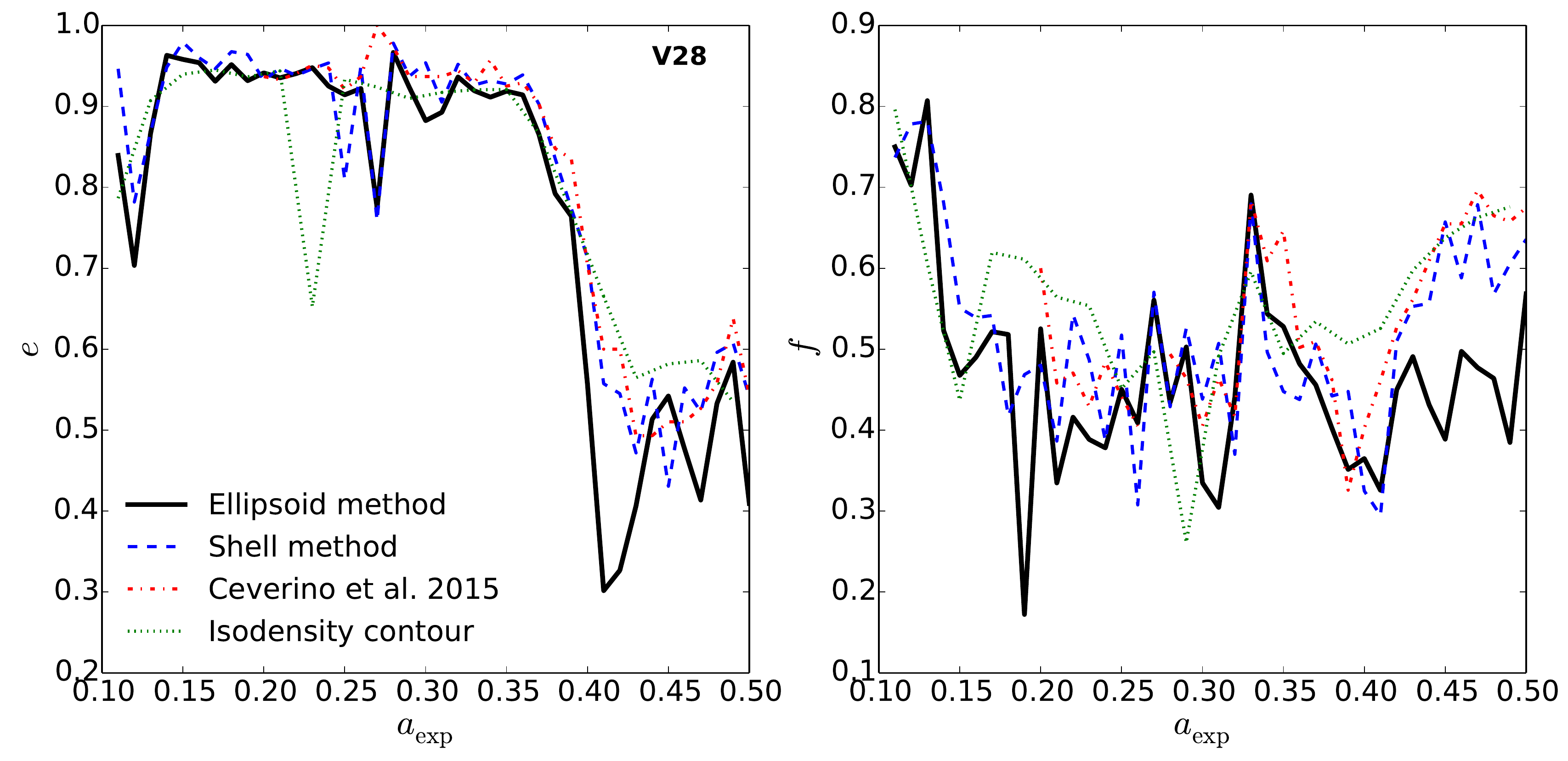}
\caption{\label{fig:ceverino_comparison} Elongation (left-hand panel) and flattening (right-hand panel) as a function of the expansion factor for the stellar component of V28. Our fiducial algorithm is shown as a black solid line and refers to the ellipsoid method, while the blue dashed line shows the results for the ellipsoidal shell method. Measurements from \citet{Ceverino+15} are shown with a red dashed line, and the ellipsoid fit to an isodensity contour about $R_e$ is plotted as a dotted green line. All methods agree to better than $\pm0.1$ in both $e$ and $f$.}
\end{figure*}

In order to test the robustness of our method to measure shape, we compare the  elongation and flattening of the stellar component of V28 as measured by a number of different algorithms.

\smallskip
One independent method that does not involve the shape tensor is to fit an ellipsoid to an isodensity contour. In particular, we compute the local particle density using an 80-particle smoothing kernel and then fit an ellipsoid to all the particles with density in the range $\rho_*\pm \delta\rho_*$.  The values of $\rho_*$ and $\delta\rho_*$ are chosen to be the median and standard deviation of the density in a shell of radius $R_e$ and thickness of 200 pc.

\smallskip
In Fig. \ref{fig:ceverino_comparison} we plot the measurements of $e$ (left-hand panel) and $f$ (right-hand panel) for V28 as it evolves in time. We show the results of our algorithm for a full ellipsoid and for an ellipsoidal shell, and compare with the fit to the isodensity contour as described above. We also show a comparison with the measurements of \citet{Ceverino+15}, which compute the shape from the inertia tensor of the 3D isodensity surface that intersects the major axis at a radius $R_{\rm e}$. The surface is selected to have a volumetric stellar density $\rho_s+\sigma_s$, where $\rho_s$ and $\sigma_s$ are the average density and its standard deviation computed for two opposite points located along the major axis at the intersection with the sphere of radius $R_{\rm e}$.

\smallskip
At most snapshots, there is agreement between all methods better than $\pm0.1$ in both $e$ and $f$.
At specific snapshots (such as at $a_{\rm exp}=0.27$), there are larger discrepancies, corresponding to large deviations from an ellipsoidal shape due to a major merger. In other cases (such as at $a_{\rm exp}=0.41$), the full ellipsoid shows a lower ellipticity or flattening than the ellipsoidal shell, at the level of $-0.2$, indicating an increasing elongation with radius between the centre and $R_{\rm e}$. We also note that after $a=0.4$ $f$ and $e$ computed for a full ellipsoid are systematically lower than those computed for an ellipsoidal shell. This indicates that the values of $f$ and $e$ are lower at radii smaller than $R_e$.

\smallskip
Unless stated otherwise, we adopt for our analysis in this paper the shape as computed by our algorithm from the shape tensor for the full ellipsoid, with the semi-major axis set to equal to $R_{\rm e}$.

\section{Shapes in Pictures}\label{sec:shape_in_pictures} 

Before we study the properties of our whole simulations suite, we focus on three example galaxies at single snapshots. These are  V7, the most massive galaxy at $z=2$, and V4 and V28 (also shown in \citet{Ceverino+15}), two prototypes of small galaxies in our sample. The shape is computed for the ellipsoid within $R_{\rm e}$ as described in Section \ref{sec:measuring_shape}, for each of the components of DM, stars and gas.

\smallskip
Fig. \ref{fig:projection_vela_prolate} shows the projections of the best-fitting ellipsoids on top of the stellar, DM and gas surface densities for V7, V4 and V28 at $z=5, 4$ and $3.5$, respectively.  The surface densities are computed projecting each component along the eigenvectors of the shape tensor of the stars. The thickness of the slice is set to be equal to the smaller eigenvalue of the stars, $\pm c_*$. 

\smallskip
From the figure it is evident that the stellar and DM components are elongated. In all three cases $c\sim b\ll a$, implying a prolate shape. In particular, we have for V7, V4, V28 respectively a stellar shape of $(e_*,f_*)=(0.93,0.55),(0.94,0.41),(0.89,0.53)$, and a dark-matter shape of $(e_{\rm DM},f_{\rm DM})=(0.91,0.6),(0.88,0.46),(0.85,0.57)$.

\smallskip
Our algorithm classifies the gas in all the three galaxies as prolate [i.e. we measure $(e_{\rm gas},f_{\rm gas})=(0.42,0.84)$ for V7, $(e_{\rm gas},f_{\rm gas})=(0.89,0.88)$  for V4 and $(e_{\rm gas},f_{\rm gas})=(0.87,0.95)$  for V28] even though its spatial distribution is very clumpy and does not have a well-defined shape.

\smallskip
Fig. \ref{fig:projection_vela_oblate} shows the shape of the different components at low redshift, $z\sim1$. Our shape algorithm gives for V7 and V4 a stellar shape of $(e_*,f_*)=(0.22,0.88),(0.37,0.72)$,  and a dark-matter shape of $(e_{\rm DM},f_{\rm DM})=(0.17,0.56),(0.32,0.36)$, which means that the stars and the DMin these two galaxies at that time are oblate ellipsoids. On the other hand, for V28 the systems are closer to being triaxial, with
$(e_*,f_*)=(0.33,0.45)$, $(e_{\rm DM},f_{\rm DM})=(0.31,0.35)$. In all simulations the gas, instead, settles into a thin disk. We note that at that time the central parts (i.e. the innermost 1 kpc) of V7 and V4 are already dominated by the baryons, while in V28 the central region is still dominated by the DM.

\smallskip
Fig. \ref{fig:ev_shape_individual}  shows the evolution tracks of the shape in the $e-f$ plane for V7, V4 and V28. Points are colour-coded by $M_{\rm DM}/M_b$ computed in the innermost 1 kpc. We see that at high redshifts both the stars and the DM are elongated, prolate or highly triaxial. Once the baryonic core becomes self-gravitating, the systems become rounder and oblate. The evolution of the shape of the gas is more complex. At high redshift, it is highly triaxial, and rather noisy, reflecting the strong deviations from ellipsoidal symmetry of the instreaming gas and the very perturbed discs. Later, and especially in the baryon-dominated phase, the gas settles to a well-defined disc.

\begin{figure*}
\begin{center}
\includegraphics[scale=0.5]{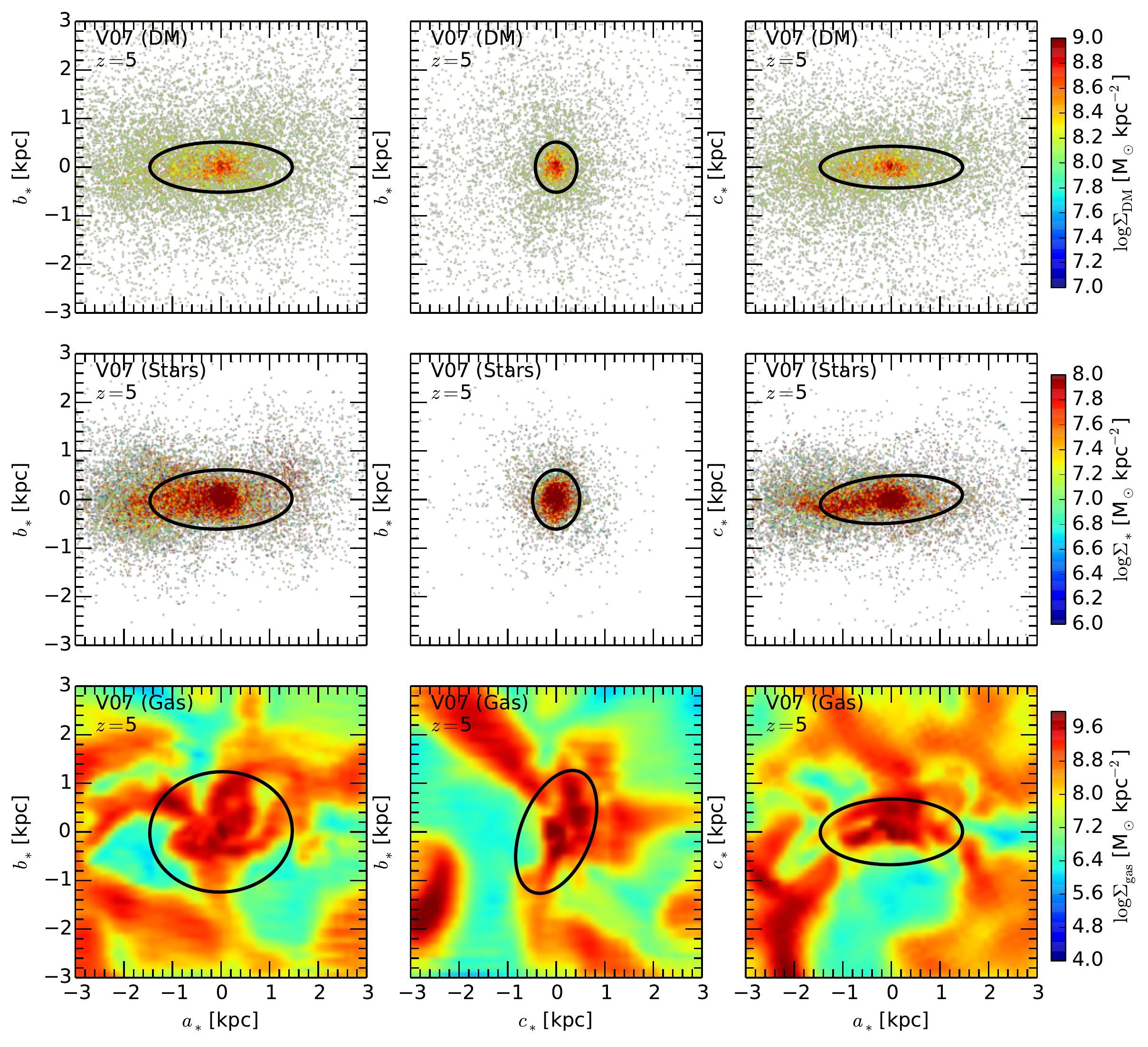}
\includegraphics[scale=0.5]
{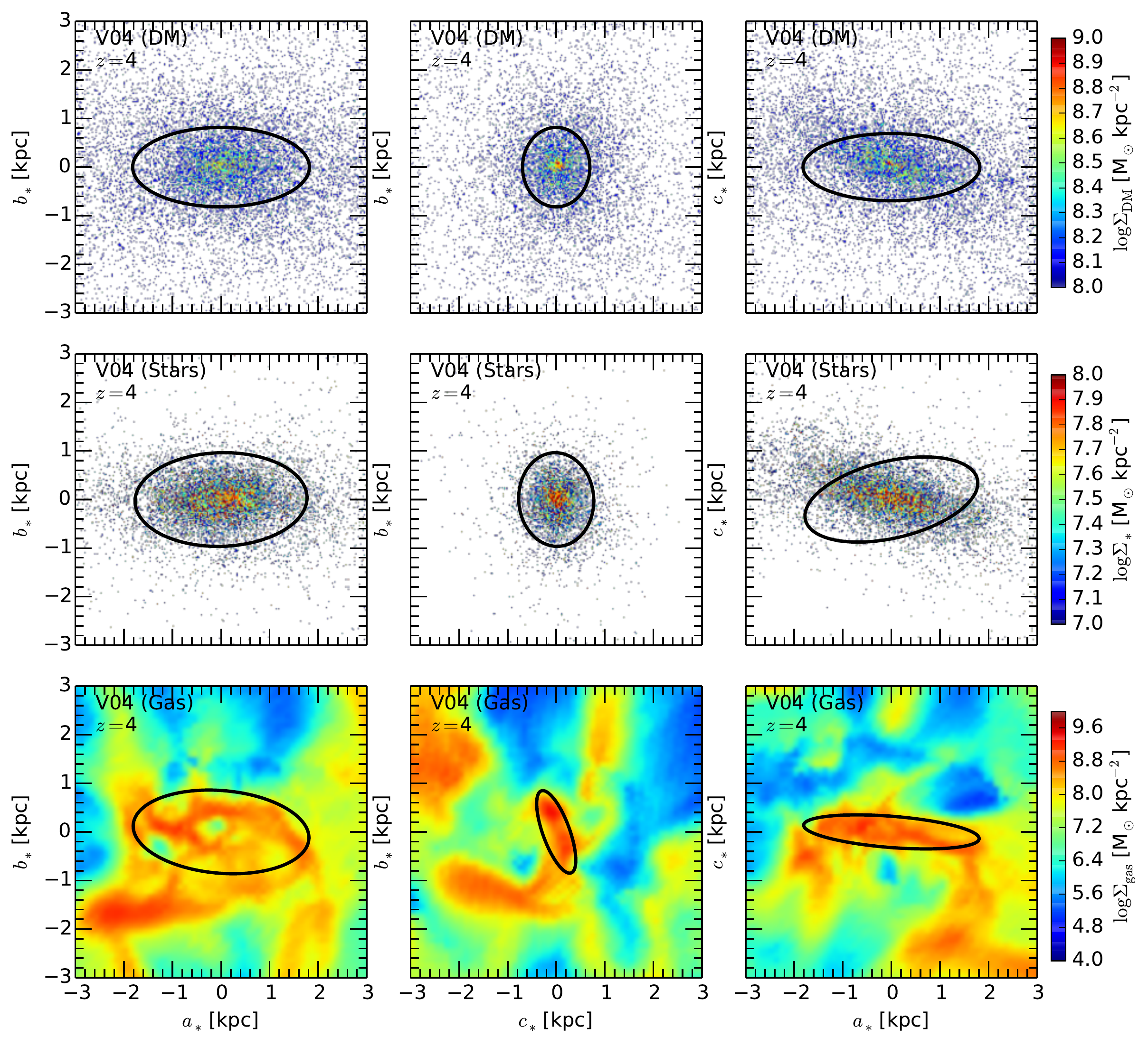}
\end{center}
\caption{DM, stars and gas surface density projections along the eigenvectors of the shape tensor of the stellar component for V7 (top) at $z\sim5$ and V4 (bottom) $z\sim4$. The thickness of the slice is $c_*$. The two-dimensional projections of the best-fitting ellipsoid computed for each component within $R_{\rm e}$ are shown in each panel. The plots depict what happens in the DM-dominated phase, when the stellar and DM systems are  prolate. The gas system in V7 is oblate, and in V4 it is triaxial, with the axes deviating from the stellar axes by 57, 53, 54 deg.}
\label{fig:projection_vela_prolate}
\end{figure*}

\begin{figure*}
\begin{center}
\includegraphics[scale=0.5]
{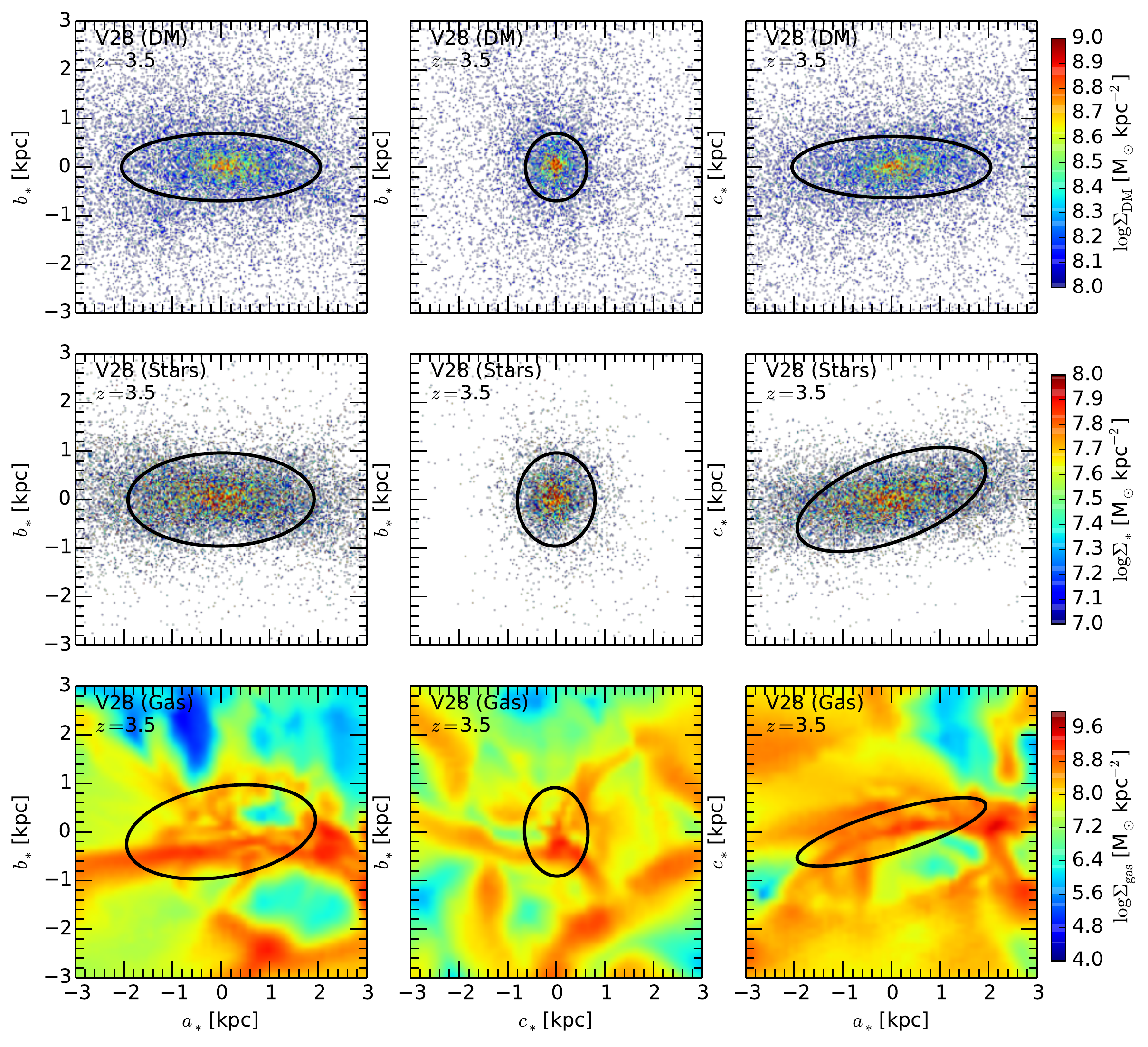}
\end{center}
\contcaption{DM, stars and gas surface density projections along the eigenvectors of the shape tensor of the stellar component for V28 at $z\sim3.5$ in the DM-dominated phase.}
\end{figure*}

\begin{figure*}
\begin{center}
\includegraphics[scale=0.5]{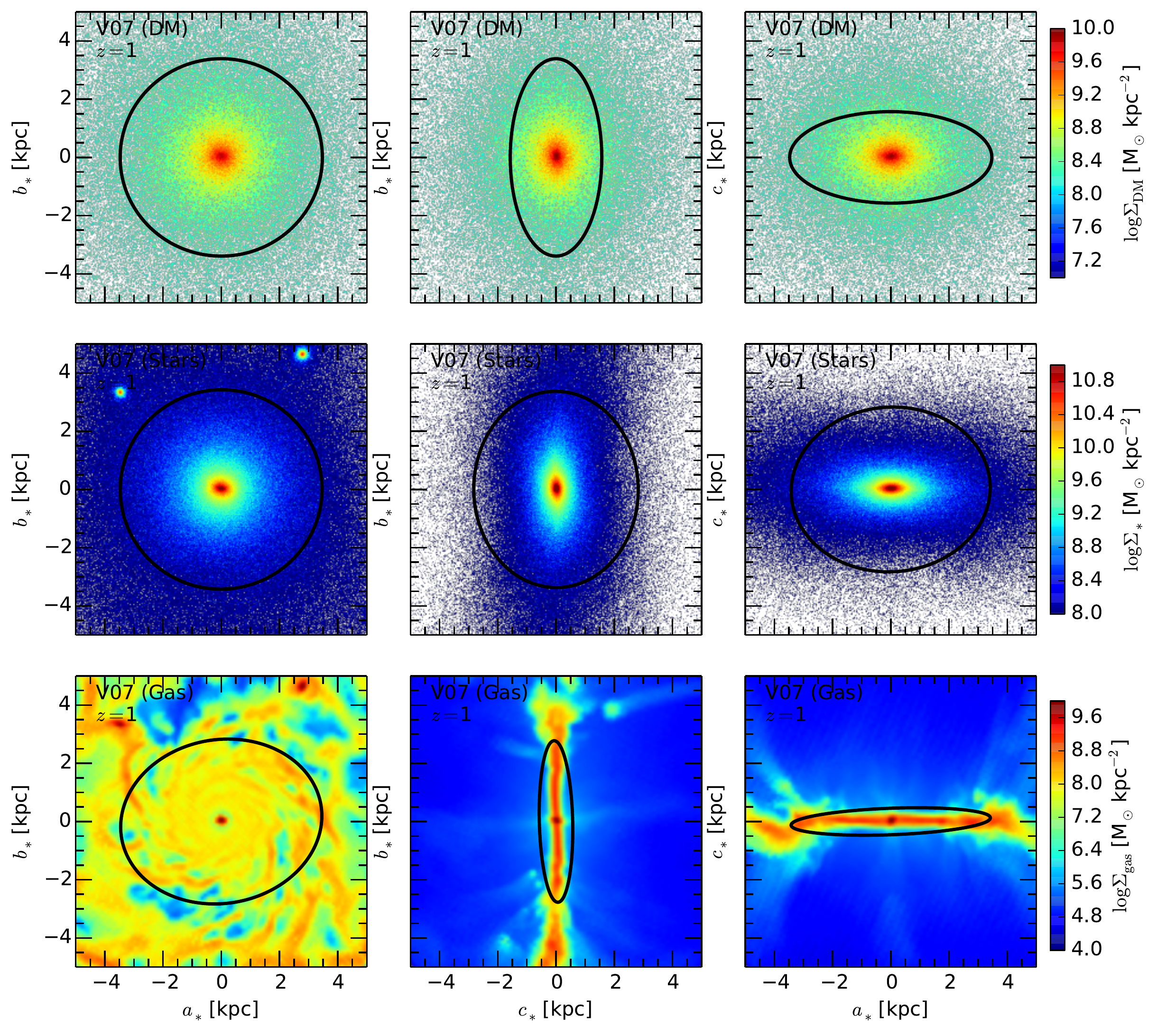}
\includegraphics[scale=0.5]
{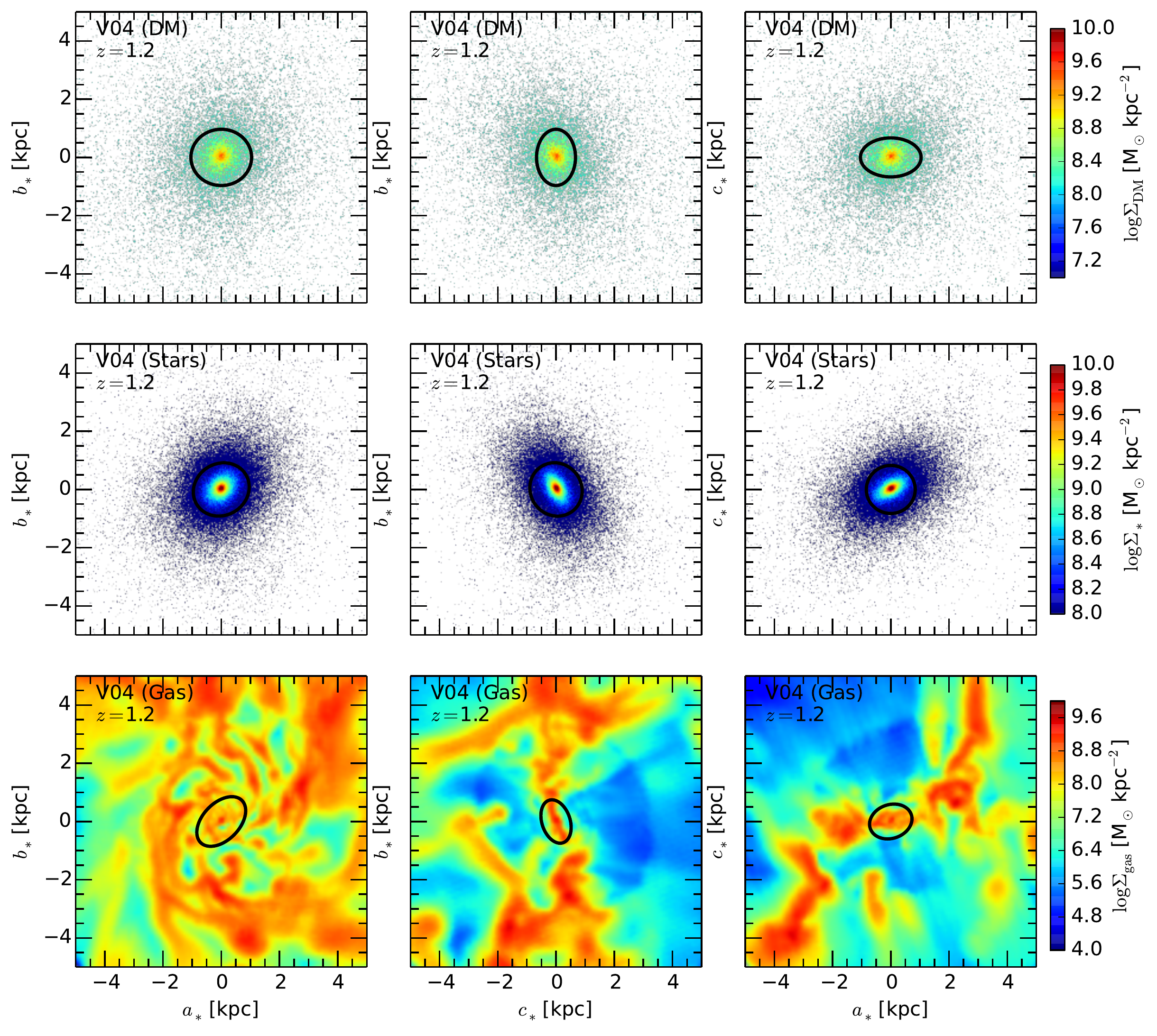}
\end{center}
\caption{DM, stars and gas surface density projections along the eigenvectors of the shape tensor of the stellar component for V7 (top) at $z=1$and V4 (bottom) at $z=1.2$. The thickness of the slice is $c_*$. The two-dimensional projections of the best-fitting ellipsoid computed for each component within $R_{\rm e}$ are shown in each panel. The plots depict what happens in the baryon-dominated regime, when the stellar, gas and DM systems are oblate.}
\label{fig:projection_vela_oblate} 
\end{figure*}

\begin{figure*}
\begin{center}
\includegraphics[scale=0.5]
{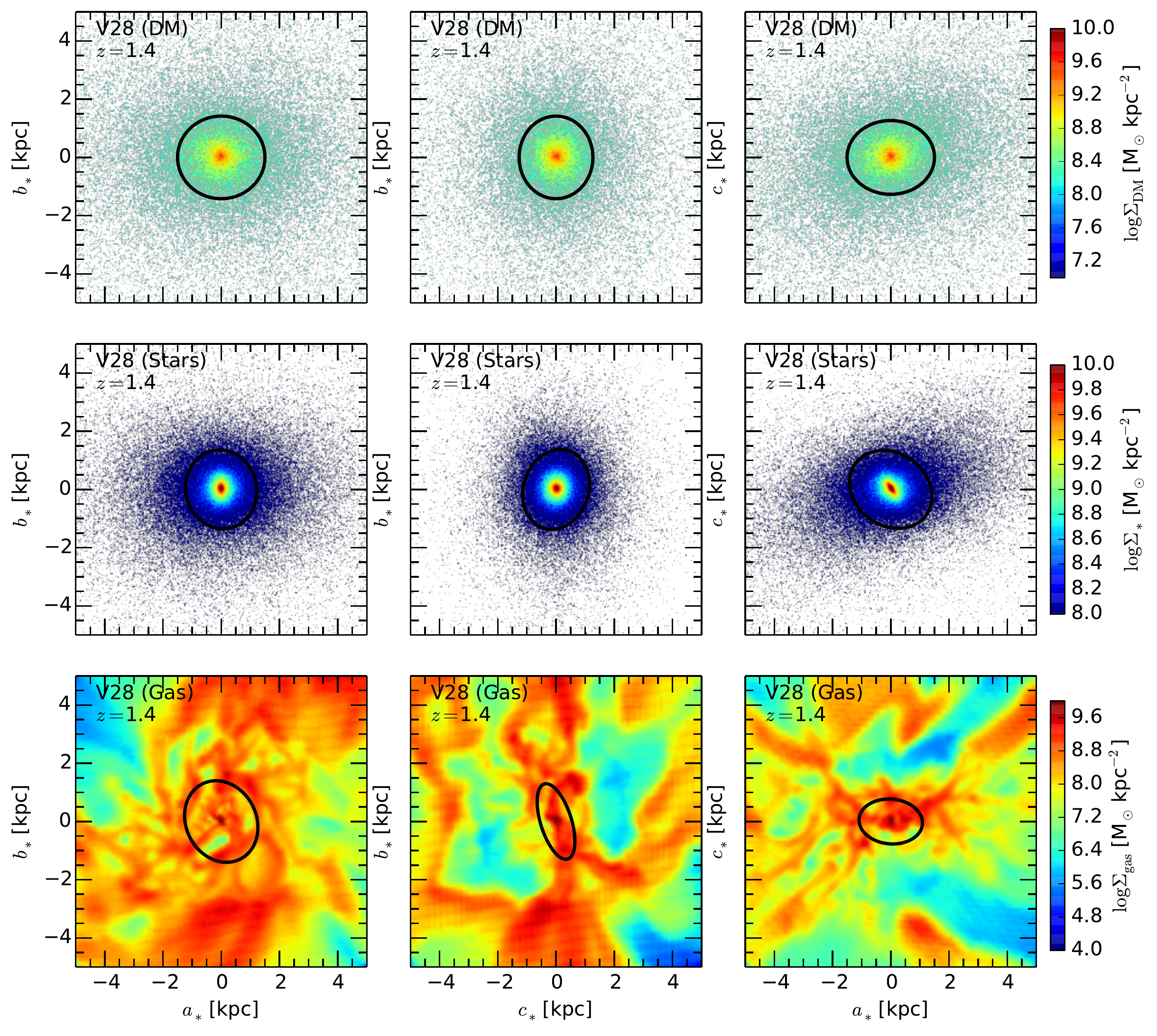}
\end{center}
\contcaption{DM, stars and gas surface density projections along the eigenvectors of the shape tensor of the stellar component for V28 at $z=1.4$ in the baryon-dominated regime.}
\end{figure*}

\section{Shape Evolution in the Whole Sample}\label{sec:evolution_of_shape} 

\begin{figure*}
\includegraphics[scale=0.35]{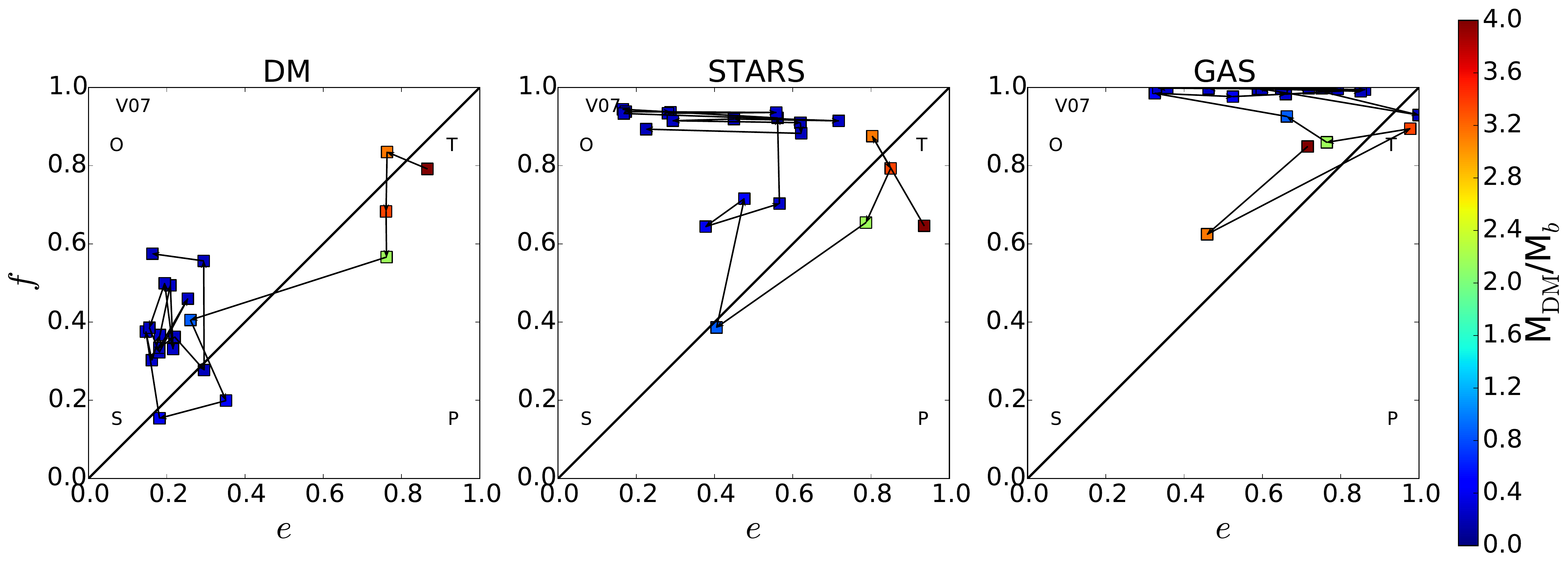}
\includegraphics[scale=0.35]{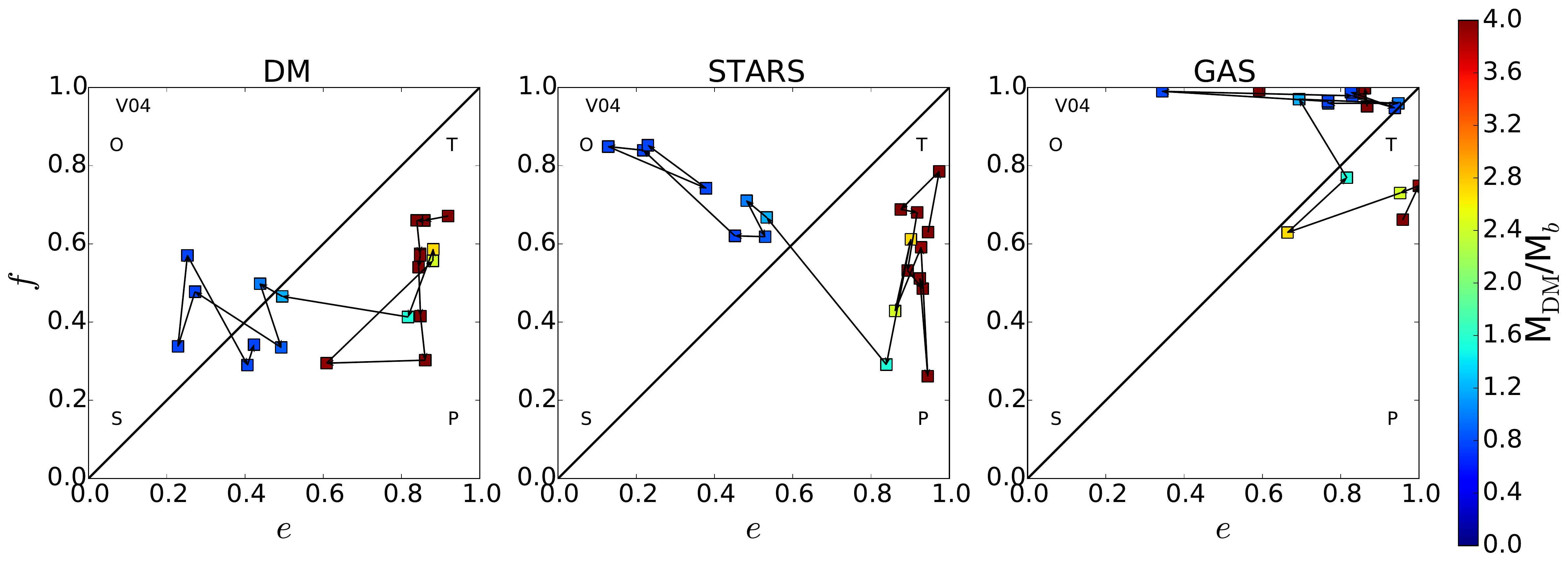}
\includegraphics[scale=0.35]{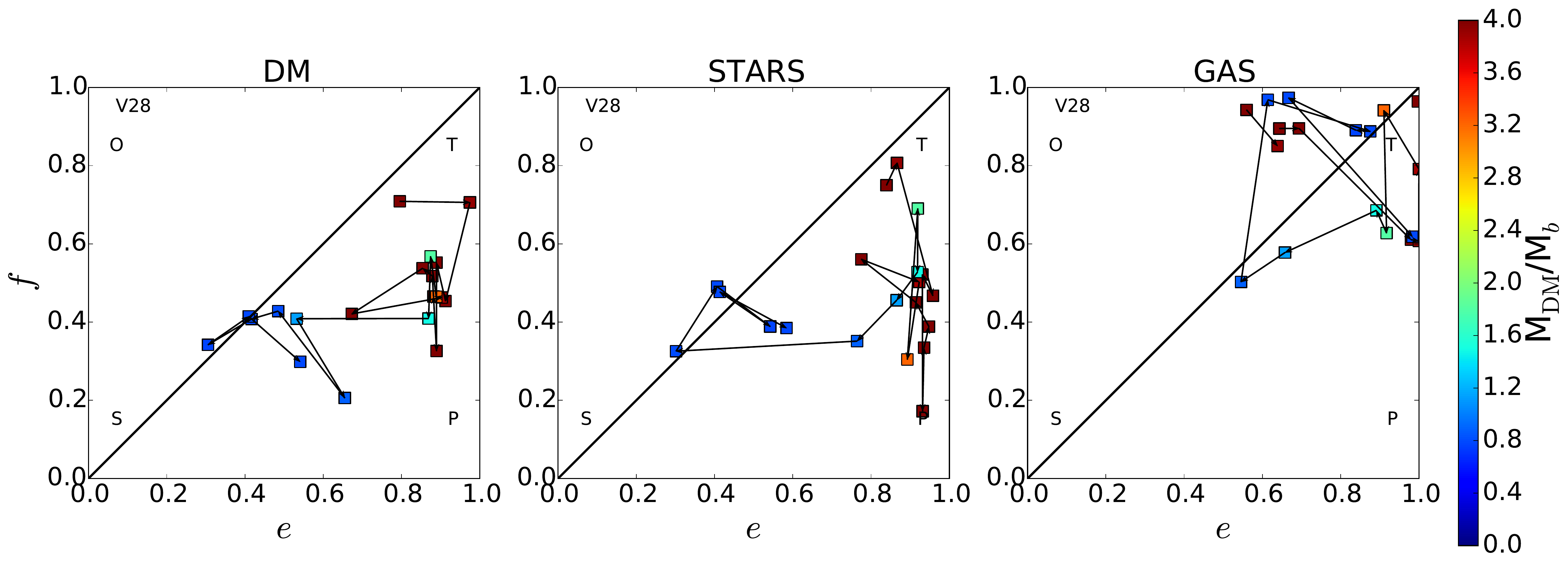}
\caption{\label{fig:ev_shape_individual} Evolution in the $e-f$ plane for V7 (top), V4 (middle) and V28 (bottom). Points are colour-coded with $M_{\rm DM}/M_b$ within 1 kpc. The systems typically evolve from high $M_{\rm DM}/M_b$ to low $M_{\rm DM}/M_b$, namely from right to left. As long as the core is DM dominated, the stars and DM are triaxial or prolate. Once the core becomes baryon dominated, they evolve into rounder-oblate systems. The gas is highly triaxial at early time, and it settles into a disc when the core is baryon dominated. The labels O, T, S and P refer to oblate, triaxial, spherical and prolate ellipsoids, as introduced in Fig. \ref{fig:ef_plane}.}
\end{figure*}

\begin{figure*}
\begin{center}
\includegraphics[scale=0.25]{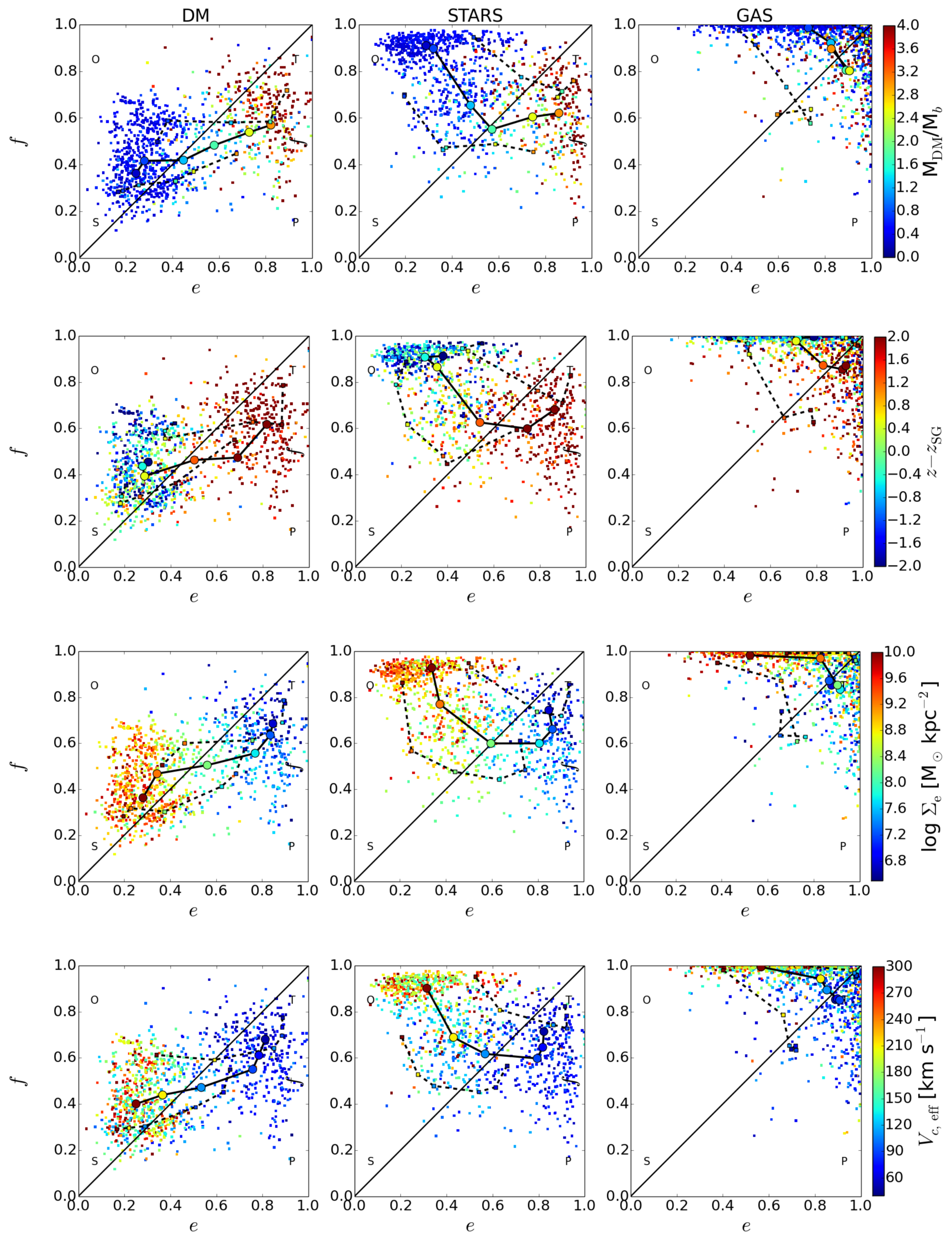}
\end{center}
\caption{Elongation versus flattening for DM (left), stars (middle) and gas (right) computed within a sphere of radius $R_{\rm e}$. The plot is colour-coded, from top to bottom, with $M_{\rm DM}/M_b$, $z-z_{\rm sg}$ (where $z_{\rm sg}$ is the redshift at which the innermost 1 kpc becomes dominated by baryons), the stellar surface density within $R_{\rm e}$ ($\Sigma_{*}$) and the circular velocity within $R_{\rm e}$ ($V_{\rm c,eff}$). Each point represents a snapshot for any of the 34 galaxies of the sample, while solid (dashed) lines show the median (20th and 80th percentiles) in bins of the quantity in the colourbar. On average, the systems are prolate in the DM-dominated epoch, at high $z$, low stellar surface density ($\Sigma_{\rm e}\lsim 10^{8.5}$ M$)\odot$ kpc$^{-2}$) and low circular velocity ($V_{\rm c,eff}\lsim100$ km s$^{-1}$)  within the effective radius. On the other hand, in the baryon-dominated epoch the halo is oblate and the stars and gas system discy. This happens at low $z$, high $\Sigma_{\rm e}$ and high $V_{\rm c,eff}$.}
\label{fig:ev_shape} 
\end{figure*}

\smallskip
Next we study the evolution of shape in our whole sample, and try to identify its origin.
Fig. \ref{fig:ev_shape} shows the evolution of shape in the plane of $e$ and $f$ for all the snapshots of all our simulated galaxies. The three columns refer to DM, stars and gas.
The colour of the symbols in the four rows refers respectively to the DM dominance within $R_{\rm e}$, $M_{\rm DM}/M_b$, the redshift with respect to the redshift of transition to self-gravity, $z-z_{\rm sg}$, the stellar surface density within $R_{\rm e}$, $\Sigma_{\rm e}$, and the circular velocity at $R_{\rm e}$, $V_{\rm c,eff}$. In order to show the general evolutionary trend, we mark with coloured-circles connected by a solid line the median values of $e$ and $f$, computed for all the galaxies in bins of the quantity in the colourbar on the right side of each plot. We also show the 20th and 80th percentile of those quantities with small squares connected by dashed lines.

\smallskip
The median evolution track for the stars starts in the prolate/highly-triaxial regime where $(e,f)\sim(0.9,0.7)$. The track becomes rounder and it crosses the triaxial line at $(0.6,0.6)$, then it evolves into the oblate regime and eventually becomes discy at $(0.3,0.9)$. The median track of the DM component also starts prolate/triaxial at $(e,f)\sim(0.8,0.6)$, and the becomes rounder, crossing the triaxial line at $(0.5,0.5)$. It then continues to become more rounder and mildly oblate, at $(0.3,0.4)$.

\smallskip
The transition from prolate to oblate typically happens near $z=z_{\rm SG}$, namely when $M_{\rm DM}/M_{\rm b}=1$. It also typically happens when the effective central surface density becomes larger than a threshold of $\Sigma_{\rm e}\sim 10^{8.5}$ M$_\odot$ kpc$^{-2}$.

\smallskip
Recall that using the same simulations, it has been shown [Fig. 1, based on \citet{Zolotov+15}, Figs. 2-4] that the transition from DM dominance to baryon dominance in the core is associated with a dramatic wet compaction event, which also triggers central gas depletion and quenching inside-out. Thus, we learn that the transition in shape is associated with the wet compaction event.

\begin{figure*}
\centering
\includegraphics[scale=0.45]{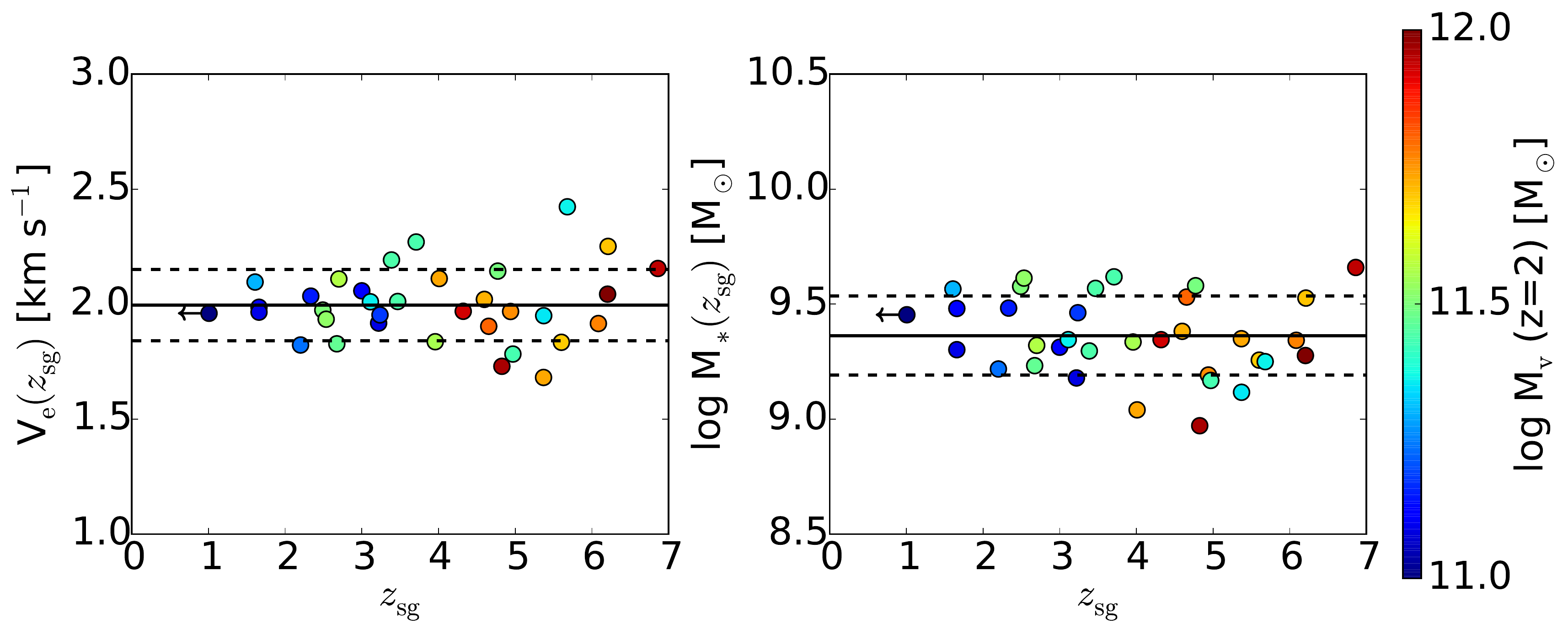}
\caption{\label{fig:self_gravity} Circular velocity (left-hand panel) and total stellar mass (right-hand panel) within the effective radius computed at the time of transition from DM dominance to baryonic dominance, $z_{\rm sg}$, versus $z_{\rm sg}$. The median and the scatter are shown by the solid and dashed lines, respectively. Points are colour-coded according to the halo virial mass at $z=2$. The escape velocity at $z_{\rm sg}$ is typically 100 km s$^{-1}$, and log $M_*\sim$  9.4, both rather independent of redshift and of the mass ranking at $z=2$. This may reflect the critical conditions below which feedback is effective in maintaining the core DM dominated.}
\end{figure*}

\smallskip
When the colour refers to the circular velocity within $R_{\rm e}$, $V_{\rm c,eff}=\sqrt{GM_{\rm e}(<R_{\rm e})/R_{\rm e}}$, the prolate-to-oblate transition is rather sharp and it tends to occur at $V_{\rm c,eff}\sim 100$ km s$^{-1}$. In Fig. \ref{fig:self_gravity}, we pick for each of our galaxies the redshift of transition from DM dominance to baryon dominance ($z_{\rm SG}$), and plot $V_{\rm c,eff}$ and $M_*$ at that time versus $z_{\rm SG}$.

\smallskip
Despite the fact that more massive galaxies (ranked at a given time) tend to compactify and become baryon dominated at an earlier time, we learn that the transition occurs for all galaxies at roughly the same escape velocity from the central region, $V_{\rm c,eff}\sim 100$ km s$^{-1}\,\pm$ 0.15 dex. A similar conclusion is valid for the stellar mass of the galaxy at $z_{\rm SG}$, with the transition at a total stellar mass of $M_*\simeq 10^{9.4}$ M$_\odot\,\pm$ 0.47 dex. We note that this velocity is in the ball park of the critical velocity below which supernova feedback is expected to drive strong outflows from the central regions \citep{Dekel+86}. This raises the suspicion that efficient feedback-driven outflow is a key for the galaxy core to remain dominated by DM, thus allowing the stellar system to keep following the elongated shape of the DM halo.

\subsection{The role of feedback}

In order to address the role of feedback in the transition to baryon dominance and the associated morphological transition to oblate shape, we appeal to a twin set of simulations, termed Gen2, which includes the same galaxies simulated with a weaker feedback. While the dominant feedback in Gen2 is supernova feedback, Gen3 incorporates also radiative-pressure feedback, from ionizing radiation [see the NoRadPre and RadPre models of \citet{Ceverino+14}].

\begin{figure*}
\includegraphics[scale=0.4]{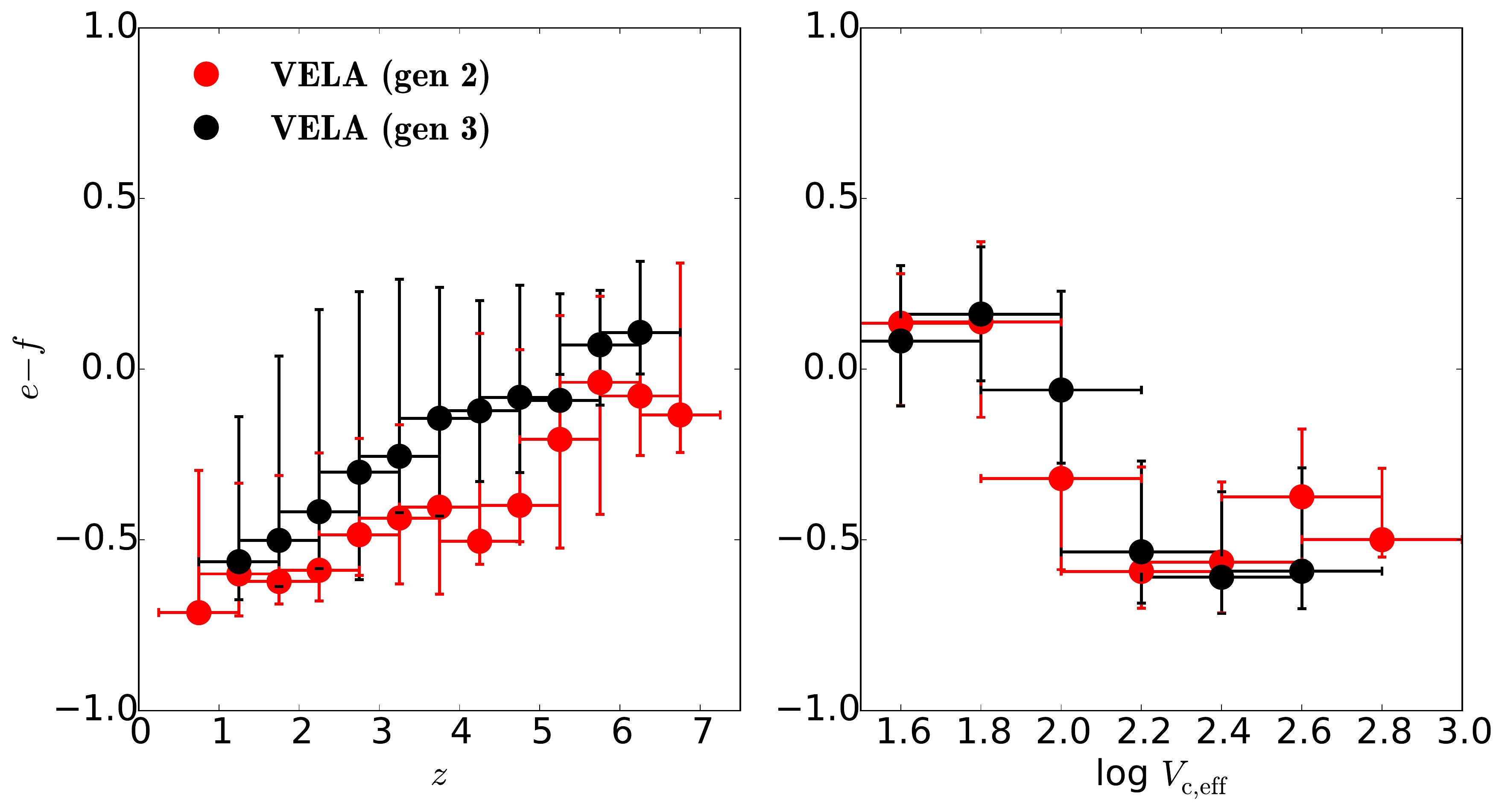}
\caption{\label{fig:effect_of_feedback} Evolution of the shape parameters $e$ and $f$ (computed within $R_{\rm e}$) for the Gen3 (black points) and Gen2 (red points) simulation suites as a function of redshift (top panels) and the circular velocity at $R_{\rm e}$ (bottom panels). The simulations with additional radiative feedback produce galaxies that are, on average, less oblate at any given time. This can be explained by the radiative feedback keeping the core DM-dominated for longer times and thus delaying the transition from prolate to oblate.}
\end{figure*}

\smallskip
The left-hand panel of Fig. \ref{fig:effect_of_feedback} compares the evolution of shape in the Gen2 and Gen3 simulations. Plotted is the quantity $e-f$, which is an effective measure of prolateness (high values of $e-f>0$) versus oblateness (low values of $e-f<0$). One can see that galaxies in the simulations of stronger feedback tend to have higher values of $e-f$, which implies that these galaxies are on average less oblate at any given time. This can be explained by the stronger feedback ejecting gas from the centre and thus keeping the core DM dominated for longer times, and postponing the transition from prolate to oblate to a later time.

\smallskip
If feedback is indeed important, one may expect that the key parameter regulating the morphology would be the escape velocity from the core. The right-hand panel of Fig. \ref{fig:effect_of_feedback}  shows $e-f$ as a function of $V_{\rm c,eff}$. One can see that there is no significant difference between the Gen2 and Gen3 galaxies, despite the different feedback strength. Indeed, there is a distinct transition from prolate to oblate once the velocity becomes larger than $\sim 100$ km s$^{-1}$, similar to what we saw in Fig. \ref{fig:self_gravity}, regardless of the feedback prescription adopted.

\section{Alignments}\label{sec:alignment}

\begin{figure*}
\includegraphics[scale=0.35]{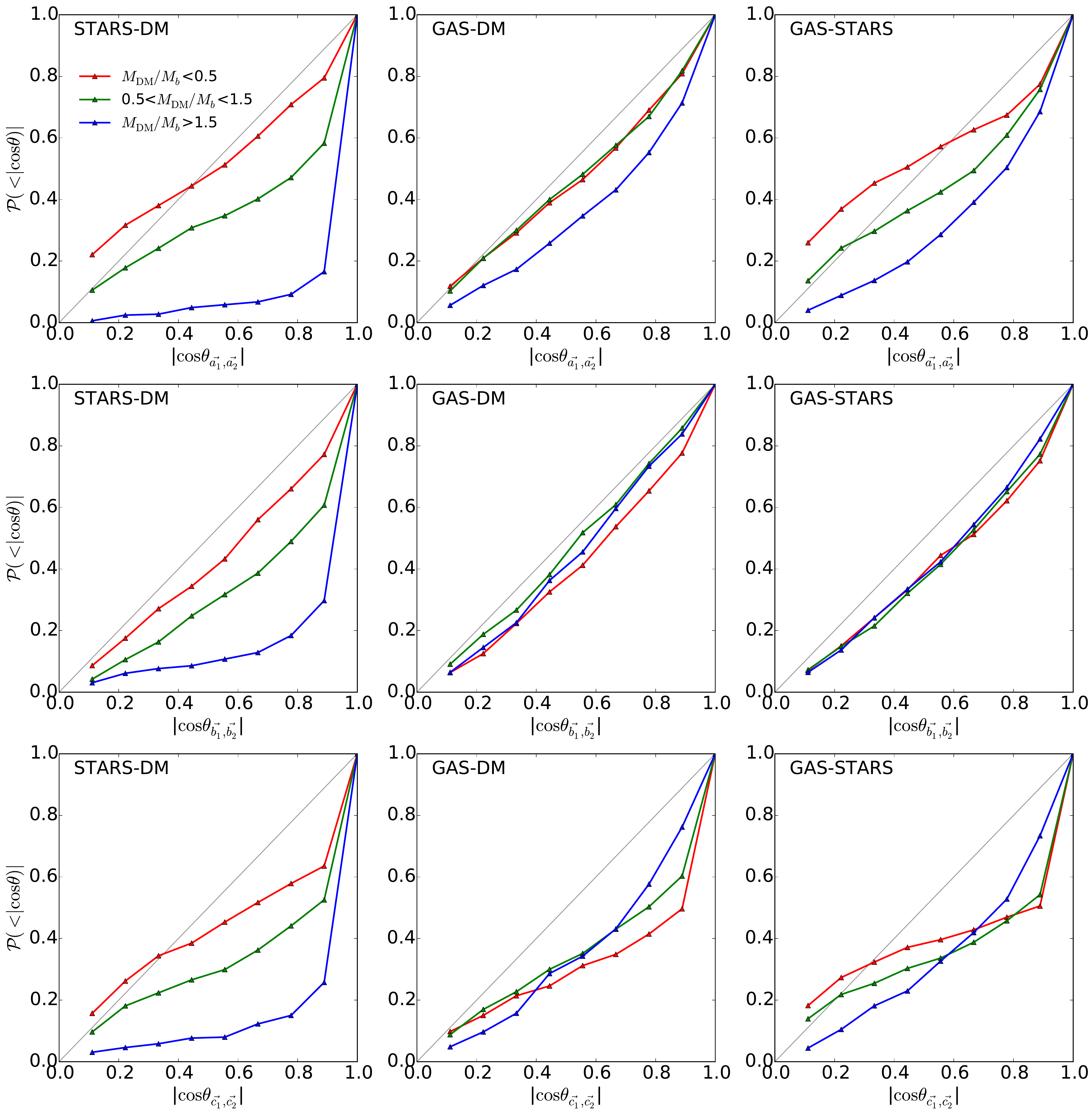}
\caption{\label{fig:alignments} Cumulative probability distributions of $|\cos\theta|$ in bins of $M_{\rm DM}/M_b$, where $\theta$ is the angle between the eigenvectors of the $\mathcal{S}$ of the different components. From top to bottom: relative orientation between the major, intermediate and minor axes. For left to right: relative orientation between stars and DM, gas and stars and gas and DM. The grey line in each panel shows the uniform cumulative distribution function. DMand stars are very well aligned with each other in the DM-dominated epoch (i.e. $M_{\rm DM/M_b}>1.5$), which supports a causal connection or a common origin between the halo and stellar shapes. In the baryon-dominated regime, when galaxies become rounder, the alignment weakens.}
\end{figure*}

In order to quantify the alignment among the different components of DM, stars and gas, we measure the angle $\theta$ between each pair of eigenvectors of the shape tensor. The coloured lines in Fig. \ref{fig:alignments} show the cumulative probability distribution functions (PDFs) of $|\cos\theta|$ in bins of $M_{\rm DM}/M_b$. In the DM-dominated epoch (i.e. $M_{\rm DM}/M_b>1.5$), stars and DM are very well aligned with each other, with the median at $\cos \theta \simeq 0.95$ for each of the pairs of major, intermediate and minor semi-axes. This is consistent with a causal effect between the halo and stellar shapes, as well as with a common origin for these shapes. On the other hand, in the baryon-dominated regime, the relative orientations for stars and DM are close to random, with a median $\cos\theta = 0.55-0.65$. This may reflect a real weakening of the alignment, but it may also be due to the galaxies becoming rounder in this regime, such that the eigenvectors are only poorly defined, especially for the DM component. The gas and stars show a significant but relatively weak alignment in the DM-dominated regime, with the medians $\cos\theta = 0.7-0.8$. This is affected by the amorphous shape of the instreaming gas. The minor axes of gas and stars become more aligned in the baryon-dominated regime, with the median $\cos\theta = 0.9$. This is because the systems settle to discs, with the directions of the minor axes determined by rotation.  

\begin{figure*}
\includegraphics[scale=0.35]{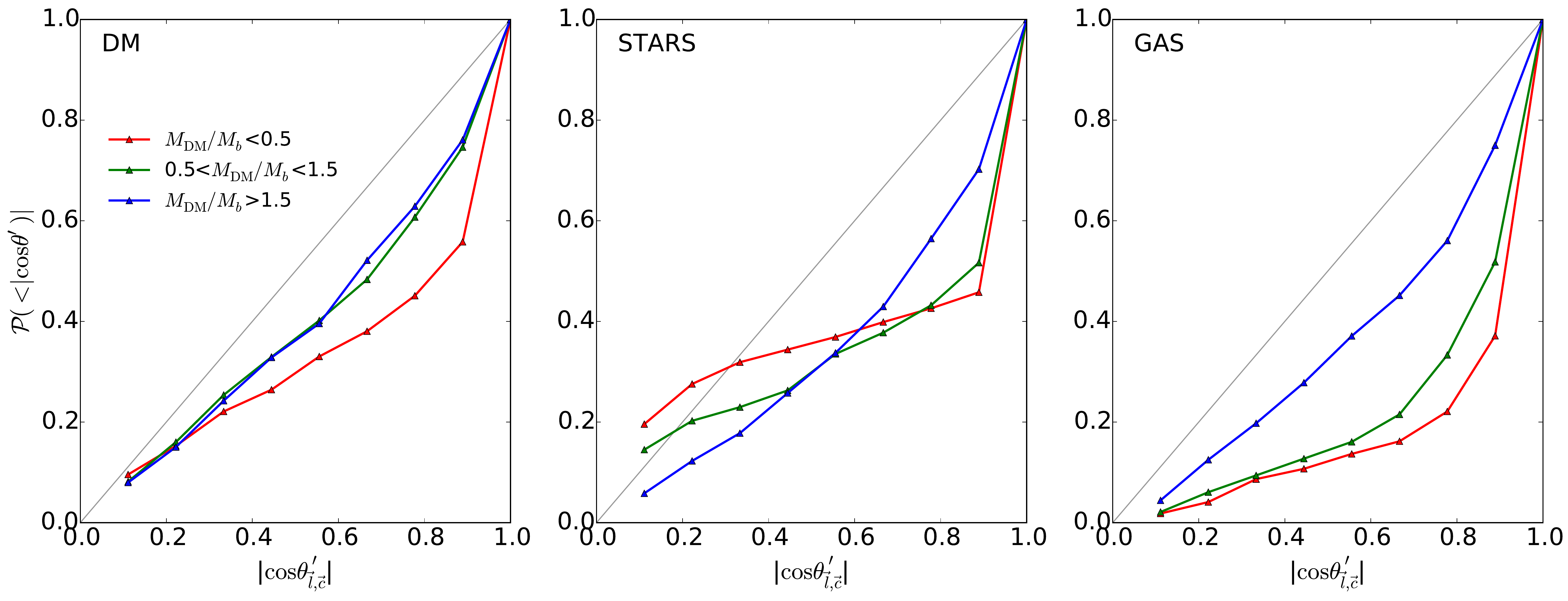}
\caption{\label{fig:alignments_spin} Cumulative probability distributions of $|\cos\theta|$ in bins of $M_{\rm DM}/M_b$, where $\theta$ is the angle between the smallest eigenvectors of $\mathcal{S}$ and the angular momentum of each component. For left to right: relative orientation between stars and DM, gas and DM and gas and stars. The grey line in each panel shows the uniform cumulative distribution function. In the baryon-dominated regime the angular momentum is aligned with the smallest eigenvector, which indicates that the shape of the stellar system is largely supported by rotation.}
\end{figure*}

\smallskip
In order to address the role of rotation, Fig. \ref{fig:alignments_spin} shows the cumulative PDF of the cosine of the angle between the smallest eigenvector, $\mathbf{c}$, and the angular momentum of each component $\mathbf{l}$, computed within the same volume, i.e. a sphere of radius $R_{\rm e}$, and find that, while in the DM-dominated regime the angular momentum is almost randomly aligned with the minor axis of the best-fitting ellipsoid; in the baryon-dominated epoch, $\mathbf{c}$ and $\mathbf{l}$ are very well aligned. 

\smallskip
This is compatible with the numerical results of \citet{Debattista+13,Debattista+15}, which studied the stability of different disc orientations within triaxial haloes and concluded that, if no gas is present, the most stable configurations are when the stellar spin is aligned with the halo minor or major axis. On the other hand, when gas is present, the disc is able to form in almost any arbitrary orientation but not perpendicularly to the intermediate axis of the halo, which is always an unstable configuration.

\smallskip
Moreover, the fact that the spins tend to align with the minor axis of the systems at later times, when the systems tend to become oblate,  is an indication  for a higher level of rotation support in the stellar system, and to a certain extent also in the DM halo, although the main support for the halo continues to be  provided by velocity dispersion.

\section{Theoretical Interpretation}\label{sec:theoretical_interpretation} 

\subsection{Origin of Halo Elongation}

In Section \ref{sec:evolution_of_shape}, we found that in the DM-dominated regime the stellar and DM systems are elongated -- triaxial and prolate (and this is true to some extent for the gas as well).
This is consistent with $N$-body simulations of DM only \citep[e.g.][]{Allgood+06}, which show that the core of a typical DM halo, soon after its assembly, tends to be prolate or triaxial.

\smallskip
This plausibly is due to the fact that haloes assemble by motions (mergers and smoother streaming) along a dominant filament of the cosmic web \citep{Dekel+06,Dekel+09}, such that the preferred direction of the filament determines the axis of largest velocity dispersion, and thus the major axis of the velocity-dispersion-supported halo.
In order to verify this hypothesis, we measure the eigenvectors of $\mathcal{S}_{\rm DM}(2\,R_{\rm vir})$, the shape tensor for the DM computed in a thin shell of radius $2\,R_{\rm vir}$, and compare it with those of $\mathcal{S}_{\rm DM}(R_{\rm e})$ and $\mathcal{S}_{\rm *}(R_{\rm e})$ for the stars and the DM on the scale of the galaxy.

\begin{figure}
\centering
\includegraphics[scale=0.4]{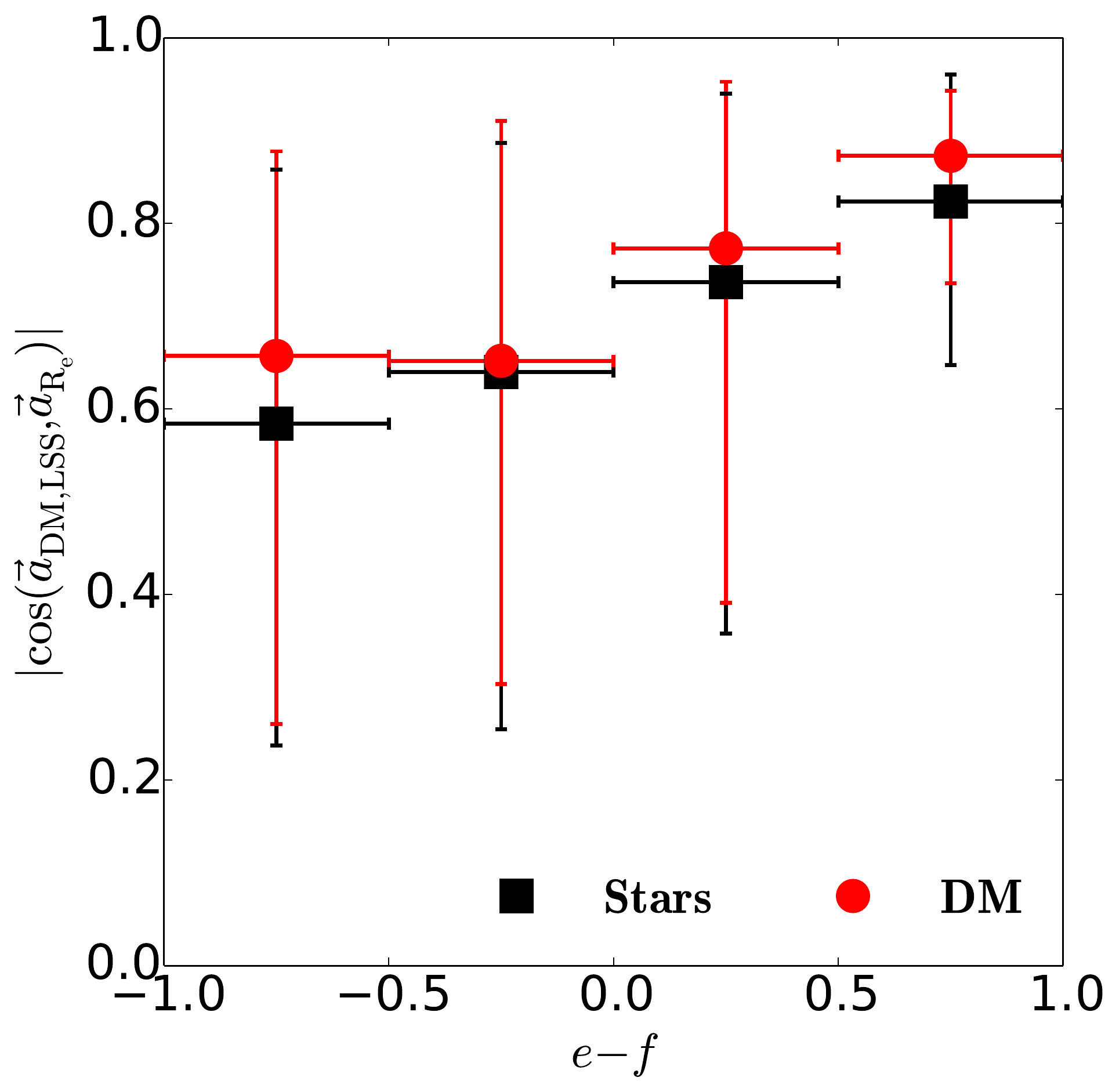}
\caption{\label{fig:large_vs_small_scales} Evolution of $|\cos(\mathbf{a}_{\rm DM, LSS},\mathbf{a}_{R_{\rm e}})|$ as a function of the stellar $e-f$. At high redshift (i.e. larger $e-f$) one can notice that there is very good alignment between the large and small scales, which suggests that the DM elongation in this epoch is dictated by the large-scale filament along which the halo is built by mergers and smoother streams.}
\end{figure}

\smallskip
Fig. \ref{fig:large_vs_small_scales}  shows the redshift evolution of $|\cos(\mathbf{a}_{\rm DM, LSS},\mathbf{a}_{R_{\rm e}})|$, where $\mathbf{a}_{\rm DM, LSS}$ is the semi-major axis of the best-fitting ellipsoid of the dark-matter component at large scales and $\mathbf{a}_{R_{\rm e}}$ is the semi-major axis of the DM (red points) or stars (black points) on small scales.
One can see that both for the DM and the stars, the alignment between the large and small scales is strong at high redshift, when $e-f>0$,  and it gradually decreases towards later times, when $e-f<0$, though some memory of the initial alignment remains. This supports the idea that the DM elongation at high redshifts, soon after the assembly, is induced by the preferred direction of the large-scale structure.  This is compatible with the results of \citet{Vera+11}, who showed using DM-only simulations that at high redshift, when haloes are fed through narrow filaments, prolateness dominates. The shape of the haloes changes to a more oblate configuration at lower redshift, when the accretion becomes more isotropic.
Analytic studies of the feeding of massive galaxies by cold streams \citep{Dekel+06}, confirmed by hydrodynamical cosmological simulations \citep{Dekel+09} and supported by comparison to observations via semi-analytic modelling \citep{Cattaneo+06}, indicate that the transition from cold inflows through narrow streams to a wide-angle inflow is expected to occur in the redshift range $z\sim1-3$, and is mass dependent.  
We indeed expect the prolateness to arise especially when one filament dominates the buildup of the halo and galaxy.
While we see that the stellar system is also aligned with the large-scale structure, we cannot yet tell whether this is because the elongation of the stellar system is a direct result of the assembly along a filament or it is induced by the elongation of the halo.

\begin{figure}
\centering
\includegraphics[scale=0.4]{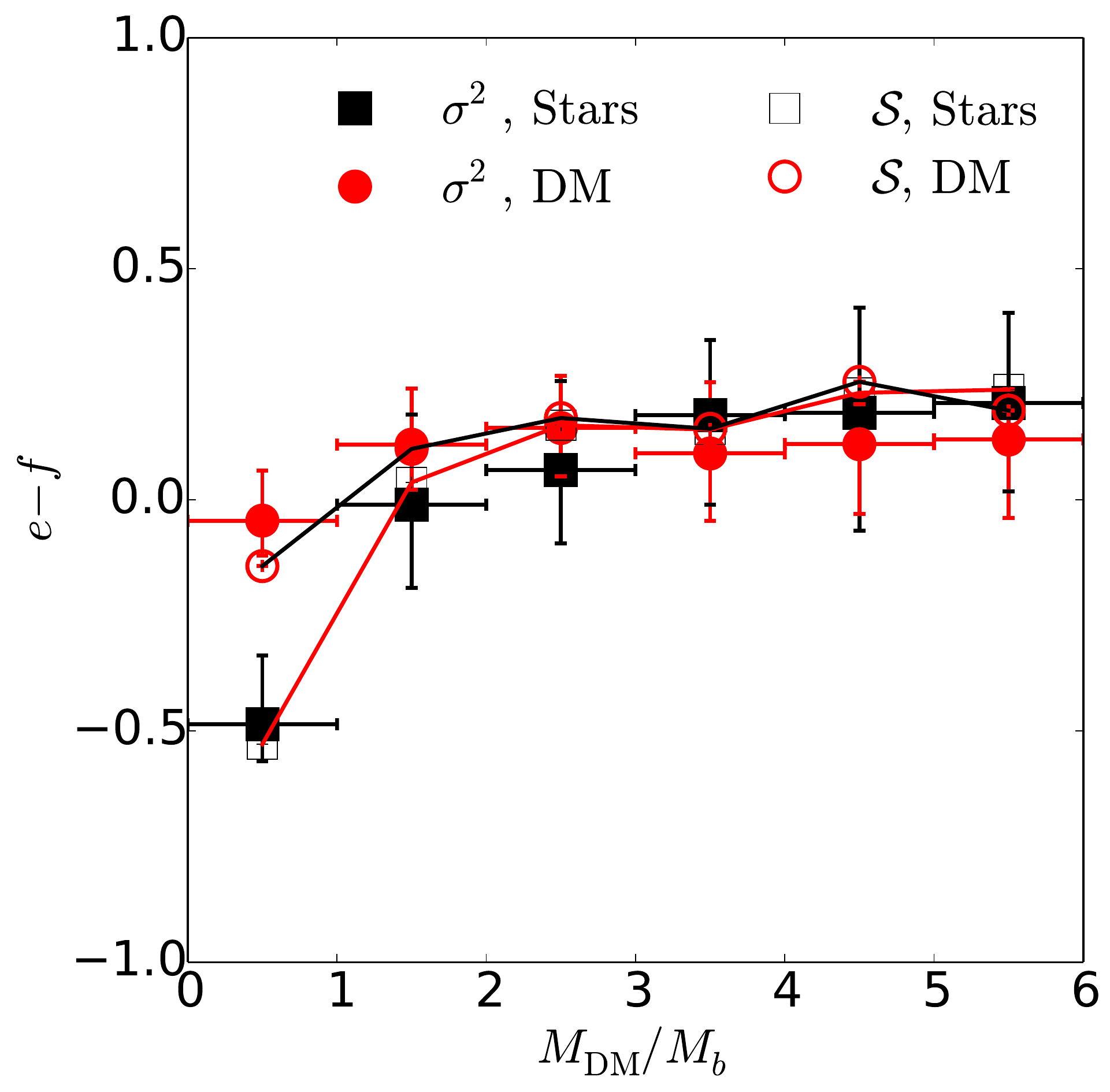}
\caption{\label{fig:ef_sigma_s} Evolution of $e-f$ as a function of $M_{\rm DM}/M_b$. Red (black) points and error bars show $e-f$ computed for $\sigma^2(\mathcal{S}$) for stars (filled symbols) and for the DM (open symbols). This shows that, for both components, the prolateness of the velocity dispersion tensor is comparable to that of the shape tensor, demonstrating that the elongated shape is supported by the anisotropic velocity dispersion.}
\end{figure}

\begin{figure*}
\centering
\includegraphics[scale=0.35]{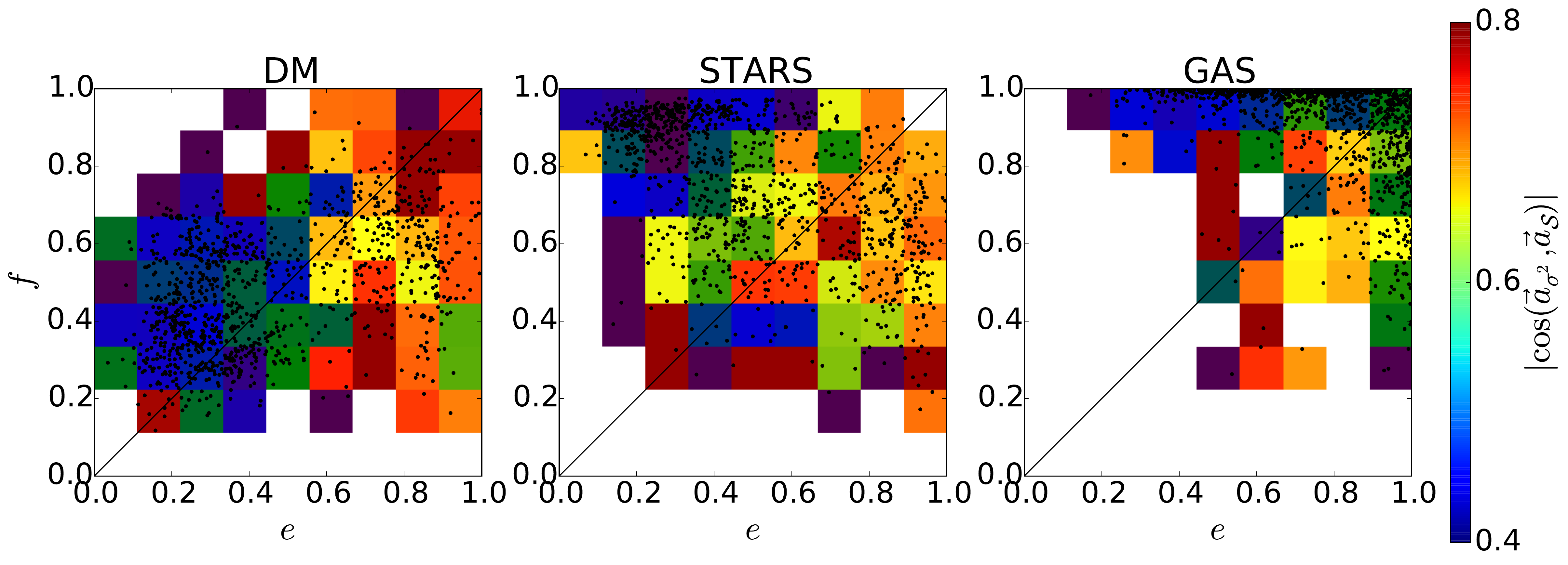}
\caption{\label{fig:shape_plot_color_direction} Black points show the elongation versus flattening for the different components (see Fig. \ref{fig:ev_shape}). The cells in the grid show the average value of the absolute value of the cosine of the angle between the major axes of the shape and velocity dispersion tensors, 
 $|\cos(\mathbf{a}_{\sigma^2},\mathbf{a}_{\mathcal{S}})|$ (computed for the stellar component) within each cell. The eigenvectors of the shape and velocity stress tensor are well-aligned when the systems are prolate, in the DM-dominated phase (where $|\cos(\mathbf{a}_{\sigma^2},\mathbf{a}_{\mathcal{S}})|$ is high). On the other hand, their relative orientation is more random at low redshift, in the baryon-dominated phase. }
\end{figure*}

\begin{figure}
\centering
\includegraphics[scale=0.4]{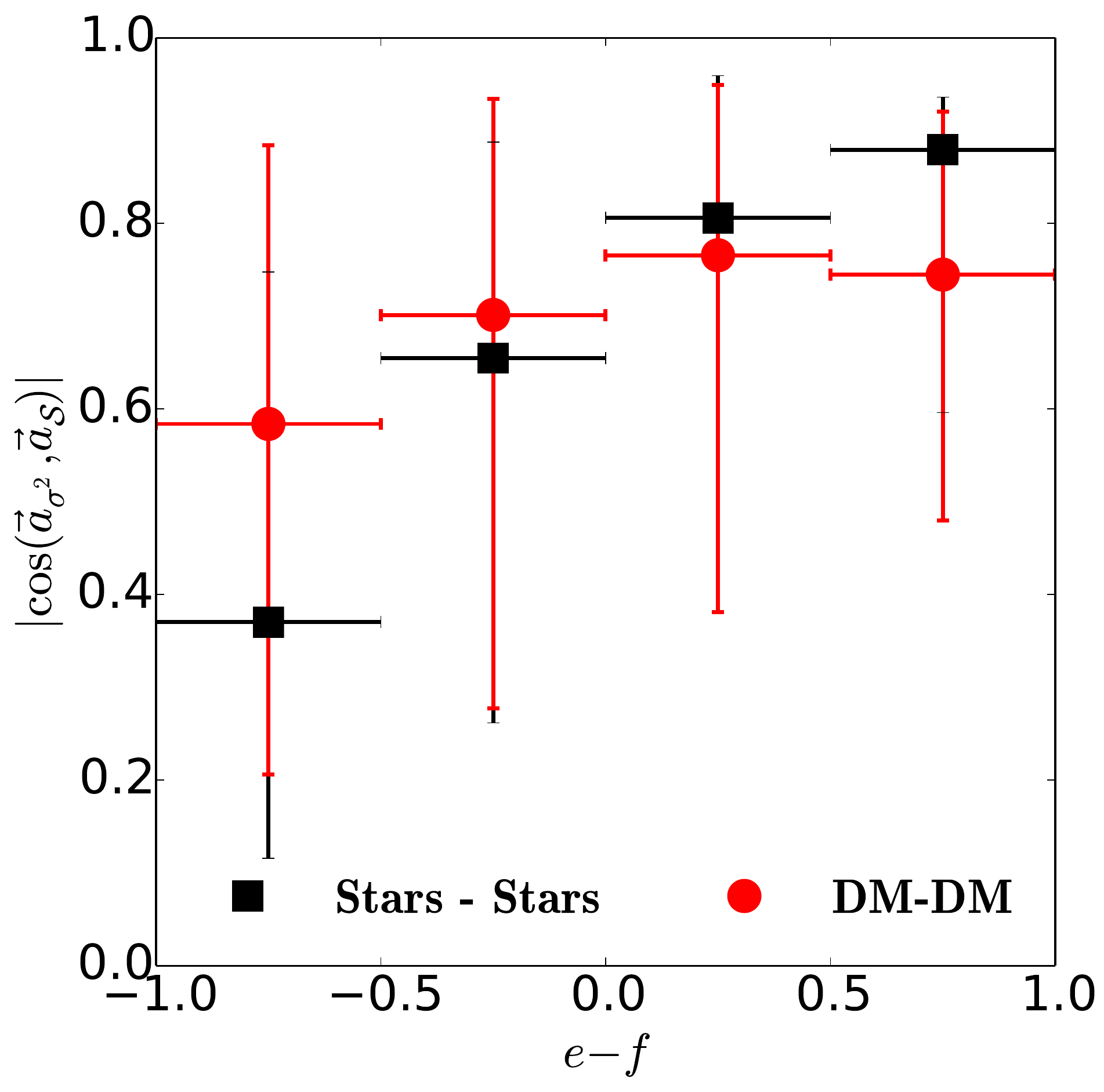}
\caption{\label{fig:sigma_s_alignment} Evolution of $|\cos(\mathbf{a}_{\sigma^2},\mathbf{a}_{\mathcal{S}})|$ as a function of the stellar $e-f$. The shape and velocity stress tensor are, on average, aligned with each other when $e-f>0$, i.e. in the DM-dominated regime.}
\end{figure}

\smallskip
In order to verify that the halo and stellar shape in the DM-dominated, prolate phase is indeed supported by anisotropic velocity dispersion, we address the alignment between the eigenvalues of the velocity dispersion tensor and the shape tensor on galactic scales. To this extent, we compute for each component, in the reference frame of its centre of mass and within a spherical region of size $R_{\rm e}$, the velocity stress tensor $\sigma^2$ defined as
\begin{equation}\label{eq:sigma_square}
\sigma_{ij}^2 = \frac{1}{M}\sum_k m_k\,(v_{k,i}-\langle v_{k,i}\rangle)\,(v_{k,j}-\langle v_{k,j}\rangle)\,\,\,,
\end{equation}
where $\langle v_{k,i}\rangle$ is the average velocity along the $i$-th component for the particle $k$, $M$ is the total mass and the index $k$ runs over all the particles in the region.
When computing $\sigma^2$, we apply the same iterative procedure adopted in the calculation of $\mathcal{S}$ (see Section \ref{sec:measuring_shape}). Based on similar arguments, the eigenvalues and eigenvectors of $\sigma^2$ can be interpreted as the squares of the semi-major axes ($a_{\sigma^2}>b_{\sigma^2}>c_{\sigma^2}$) and axis orientations ($\mathbf{a}_{\sigma^2},\,\mathbf{b}_{\sigma^2},\,\mathbf{c}_{\sigma^2}$) of the best fitting ellipsoid.
Fig. \ref{fig:ef_sigma_s} shows the evolution of $e-f$ for both the shape (black symbols) and the velocity stress tensor (red symbols) as a function of $M_{\rm DM}/M_b$. We see that the prolateness of the velocity dispersion is indeed comparable to that of the shape, both for the DM and for the stars, to within $\pm 0.1$. 
Fig. \ref{fig:shape_plot_color_direction} shows the $e-f$ plane for all components colour-coded with the average $|\cos(\mathbf{a}_{\sigma^2},\mathbf{a}_{\mathcal{S}})|$ for the stellar component. One can see that there is good alignment with the shape and velocity stress tensor when systems are prolate, while their relative orientation is more random in the baryon-dominated phase. We quantify this result in
Fig. \ref{fig:sigma_s_alignment}, which shows that the major eigenvectors of $\mathcal{S}$ and $\sigma^2$ are well aligned when the systems are prolate, both for the DM and for the stars.
At later times, towards the baryon-dominated phase, the alignment weakens,
more so for the stars. This reflects the evolution of the stellar system towards more rotation support, induced by new star formation from rotating cold gas, while the halo remains primarily velocity dispersion supported.  

\subsection{Torques by the Halo on the Stellar System}

Next we wish to find out whether the prolate DM halo when it dominates the core can induce an aligned prolate shape on the stellar system. A first crude step would be to verify the validity of the necessary condition that the amplitude of the torque exerted by the halo is sufficient for significantly tilting the stellar system.  

\smallskip
When measuring the torques exerted by DM and gas in the halo on the stellar and gaseous core, we discretize the matter distribution on large scales on a grid with a resolution of 1 kpc, while we adopt a finer grid, with resolution of 100 pc, for the material within $R_{\rm e}$.
The torque acting in each point is:
\begin{equation}
\bm{\tau} = \sum_{R<R_{\rm e}}\mathbf{r}\times\mathbf{F}\,\,\,,
\end{equation}
where $\mathbf{r}$ is the position vector relative to the centre of the galaxy and $\mathbf{F}$ is the total force acting there.
We assume that the moment of inertia about the minor axis of the ellipsoidal body within $R_{\rm e}$ is $I_c=(1/5)M_{\rm e}(a^2+b^2)$ and constant between consecutive output timesteps. Then the rate of change of the angular velocity exerted by the torque about the minor axis is $\dot\omega_c = \tau_c/I_c$.

\smallskip
During an orbital time, $t_{\rm orb} = 2\pi\,\sqrt{R_{\rm e}^3/(G\,M_{\rm e})}$, the torques are capable of rotating  the system about the minor axis by $\Theta \sim \dot \omega_c t_{orb}^2$.

\smallskip
We measure the potential effectiveness of the torques by the parameter
\begin{equation}\label{eq:alpha}
\alpha = 1-|\cos\langle\Theta\rangle|\,\,\,.
\end{equation}
If $\alpha\sim1$, the torques are potentially capable of significantly tilting the system. On the other hand, if $\alpha\ll 1$ (as long as $\Theta<\pi/2$) the torques  have a negligible effect on the orientation of the object.

\begin{figure*}
\includegraphics[scale=0.3]{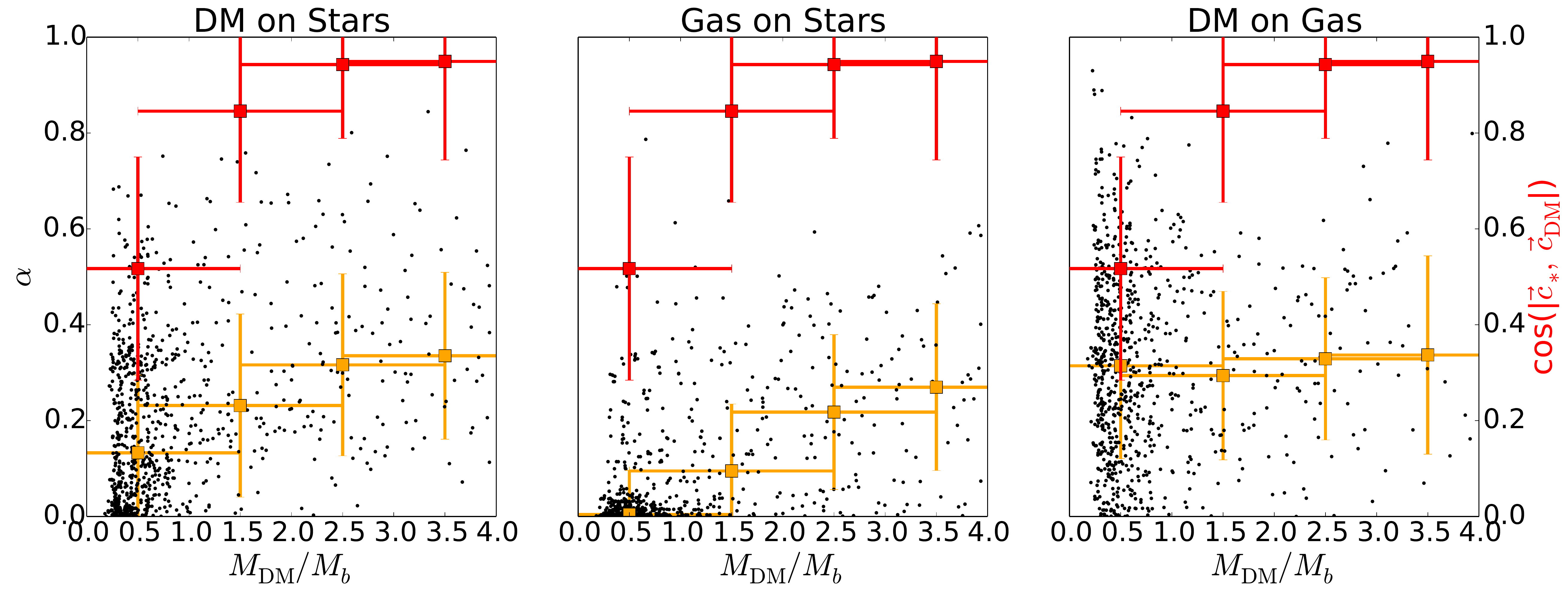}
\caption{\label{fig:torques_strength} Black points show the value of $\alpha$, which measures the effectiveness of the torques exerted by one component to the other (see equation \ref{eq:alpha}) as a function of $M_{\rm DM}/M_b$ within 1 kpc for every component and for every snapshot in our simulation suite. Orange squares and errorbars show the median $\alpha$ and its scatter in bins of$M_{\rm DM}/M_b$. The red points and errorbars, instead, show the median value of the cosine of the angle between the semi-minor axis of stars and DM as a function of $M_{\rm DM}/M_b$. The values of $\alpha\sim0.3$ in the DM-dominated era indicate that the torques exerted by the DM on the stars and gas may be sufficient for causing a significant change in the orientation of the latter, thus aligning the baryons with the DM halo. These torques become ineffective when the core becomes baryon dominated.}
\end{figure*}

\smallskip
Fig. \ref{fig:torques_strength} shows as black points the average $\alpha$ over the fine grid within the $R_e$ ellipsoid versus $M_{\rm DM}/M_b$ for all the snapshots of all the simulated galaxies. The three panels refer to the effect of DM on stars, gas on stars and DM on gas, respectively.  The orange squares refer to the median of $\alpha$ in bins of $M_{DM}/M_b$ computed within 1 kpc.  Also shown (red squares) is the median cosine of the angle between the minor axes of DM and stars in the same bins.
One can notice that the torques of DM on stars and gas on stars are stronger in the DM-dominated phase where $\alpha \sim 0.35$ they are capable of rotating the system within one orbital time by an angle as large as $\sim50$ deg. This is a hint that the torques by the halo can generate the alignment of the stellar system with it. On the other hand, when baryons dominate, the torques get progressively weaker, because the halo becomes rounder, while the stellar inertia tensor becomes larger due to the growing stellar mass fraction. The effect of the torques of DM on gas in the DM-dominated regime is comparable to that of the DM on stars, reflecting a comparable moment of inertia for gas and stars. In the baryon-dominated regime, the effect of DM on gas remains high, because the gas moment of inertia becomes low due to the low gas fraction.
The torques of gas on stars are also important in the DM-dominated phase, due to the high gas fraction then, and they become negligible in the baryon-dominated phase, where the gas fraction becomes small.

\begin{figure}
\centering
\includegraphics[scale=0.4]{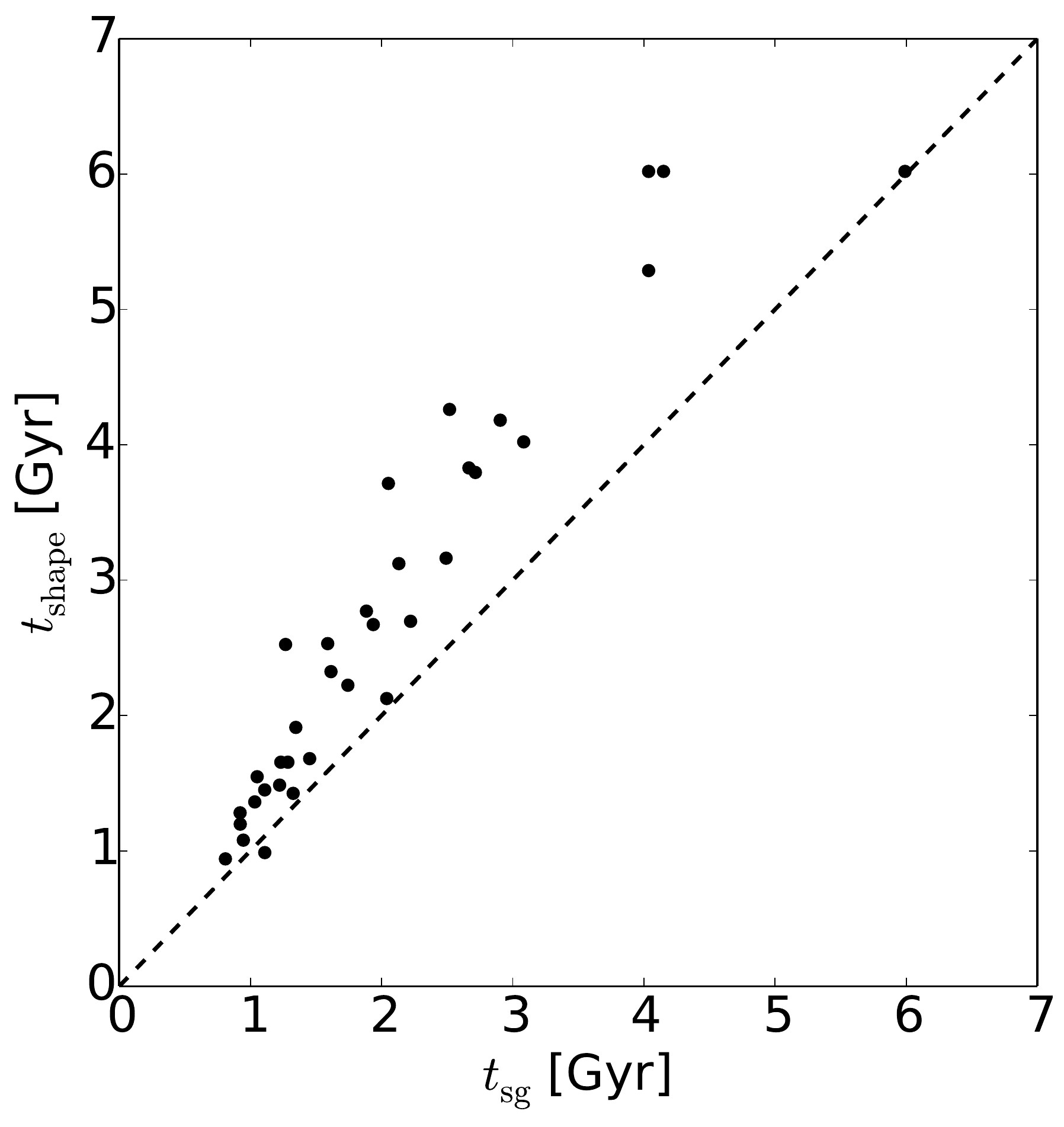}
\caption{\label{fig:scatter_plot_times} Scatter plot for all galaxies in our sample showing $t_{\rm sg}$, the time when the innermost part of a galaxy becomes self-gravitating, versus $t_{\rm shape}$, defined as the time when a system becomes strongly oblate. The black dashed line shows the locus of points that have $t_{\rm sg}=t_{\rm shape}$. For all galaxies (but one), we have $t_{\rm shape}>t_{\rm sg}$, which means that the systems become strongly oblate only after they have accumulated a large baryonic mass in the innermost regions.}
\end{figure}

\subsection{Smearing of Elongation}

Once the baryons in the core become self-gravitating, the halo and the stellar systems become rounder. This happens once the baryonic component is massive enough ($M\sim10^{9.4}$ M$_\odot$) and compact ($\Sigma_{\rm *,eff}\sim10^{9}$ M$_\odot$ kpc$^{-2}$). A central mass concentration could in principle deflect small-pericentre orbits of stars and DM particles and thus make the inner parts of the system rounder, as shown in $N$-body simulations \citep[e.g.][]{Athanassoula+05,Debattista+08}. While these simulations address a somewhat different problem of destructing a bar in the centre of a disc and are therefore not applicable quantitatively to the problem in hand, they demonstrate the potential effectiveness of a central density peak in making an elongated system supported by boxy orbits rounder. 

\smallskip
In order to understand the connection between the compaction event and the transition to an oblate system we compare two time-scales. First, we compute $t_{\rm sg}$, the time that corresponds to $z_{\rm SG}$, i.e. the epoch at which the inner part of systems becomes baryon dominated, $M_{\rm DM}/M_b<1$. Next, we define $t_{\rm shape}$ as the average time of the first four crossings of the line $e_*-f_*=-0.4$ \footnote{In the $e_*-f_*$ plane, the $e_*=f_*$ line identifies systems that are perfectly triaxial and separates oblate and prolate systems. In particular, oblate system will lie above this line and the transition is somehow arbitrary, but an intercept of $-0.4$ ensures that systems with $e_*-f_*>-0.4$ are strongly oblate}, which means that for $t>t_{\rm shape}$ the system is strongly oblate. Only three low-mass galaxies (V2, V5 and V28) do not become oblate according to this demanding criterion by the end of the simulation and in these cases we set  $t_{\rm shape}$ to be equal to the time of the latest snapshot of each of these simulations, at $z_{\rm fin}$.

\smallskip
Fig. \ref{fig:scatter_plot_times} shows $t_{\rm sg}$ versus $t_{\rm shape}$ for all galaxies in our sample. We see that in all cases (with one exception) we have $t_{\rm shape}>t_{\rm sg}$, which is an indication that the shape transition happens after the systems have accumulated a significant baryonic mass in the innermost regions.

\smallskip
Moreover, as one can see in Fig. \ref{fig:torques_strength}, the rounder the systems get, the weaker the torques become. In this situation, the elongated DM halo is unlikely to cause severe angular-momentum loss in the gas, and cannot generate yet another wet compaction.

\section{Projected Shapes}

A three-dimensional ellipsoid projected along the plane of the sky has a two-dimensional axial ratio $Q = B/A$ equal to \citep{Stark+77,Binney+85}
\begin{equation}\label{eq:shape_2D}
Q = \sqrt{\frac{(Q_1+Q_3)-\sqrt{(Q_1-Q_3)^2+Q_2^2}}{(Q_1+Q_3)+\sqrt{(Q_1-Q_3)^2+Q_2^2}}}\,\,\,,
\end{equation}
where
\begin{align}
& Q_1 = \frac{{\cos\theta}^2}{s^2} \left({\sin\phi}^2+\frac{{\cos\phi}^2}{q^2} \right)+\frac{{\sin\theta}^2}{q^2}\,\,\,, \\
& Q_2 = \frac{1}{s^2}\cos\theta\sin(2\phi)\left(1-\frac{1}{q^2}\right)\,\,\,, \\
& Q_3 = \frac{1}{s^2}\left(\frac{{\sin\phi}^2}{q^2}+{\cos\phi}^2\right)\,\,\,.
\end{align}
Here $\theta$ is the angle between the line of sight and the minor axis of the ellipsoid, and $\phi$ is the angle between the intermediate axis of the ellipsoid and the line of nodes, which is a line on the plane of the sky that is perpendicular to the minor axis.

\smallskip
Given the distribution of $Q$ for a sample of observed galaxies, one can use equation (\ref{eq:shape_2D}) to put constraints on the distributions of the 3D axial ratios based on certain assumptions concerning the shapes of these distribution functions.  For instance, \citet{Chang+13} and \citet{vanderWel+14} assumed that $T$ and the edge-on ellipticity, $1-s$, are Gaussianly distributed and concluded that from the observed distribution of $Q$ low-mass galaxies (i.e. $M_*\le10^{10}$ M$_\odot$) are more prolate at $z\gsim1$, while they are consistent with a population of oblate spheroids at the current epoch. Similar conclusions were drawn by \citet{Takeuchi+15}, who assumed, instead, that the face-on ellipticity, $1-q$,  is distributed in a lognormal fashion and that $s$ follows a Gaussian distribution.

\begin{figure*}
\includegraphics[scale=0.4]{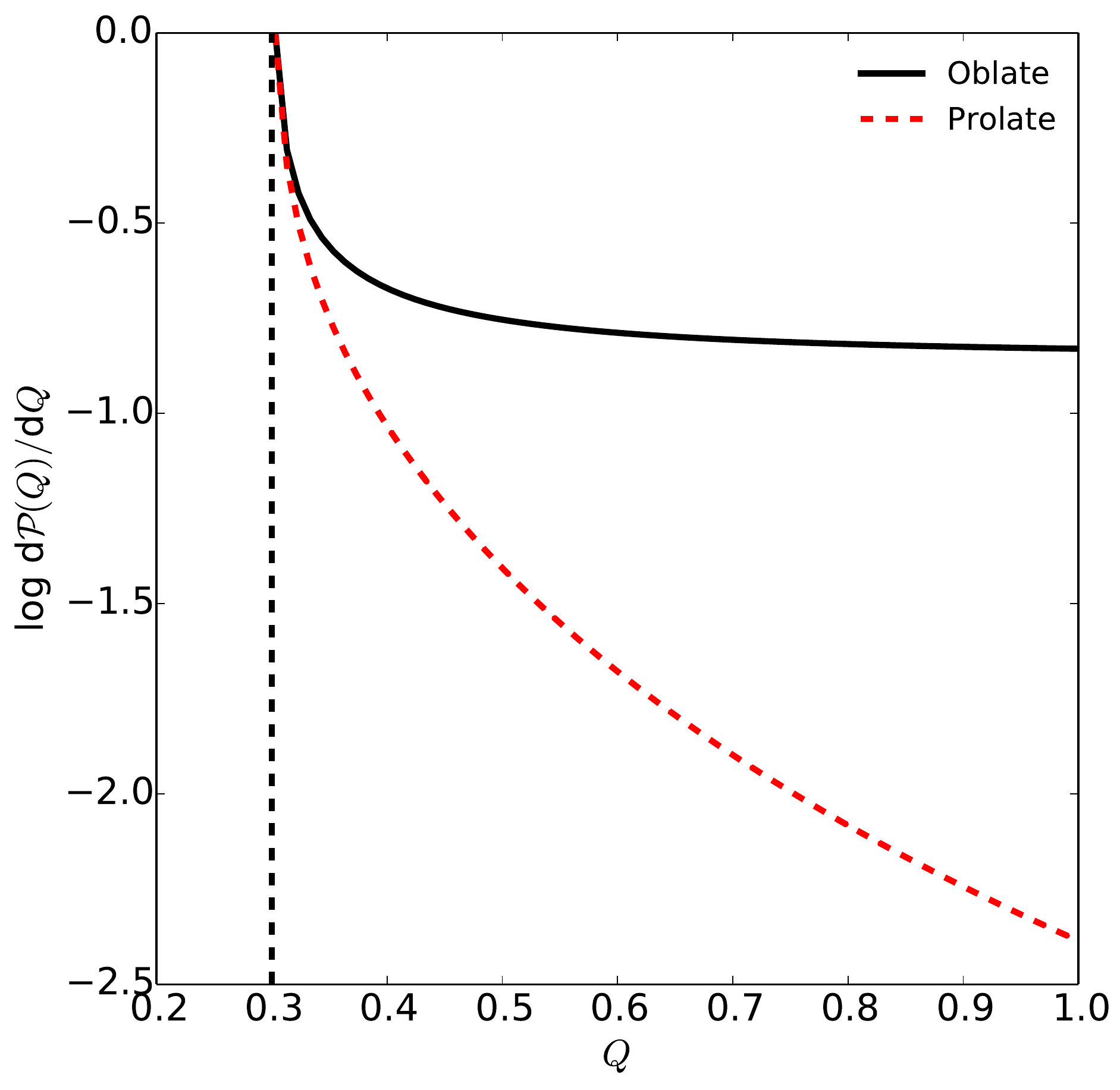}
\caption{\label{fig:distr_pure_systems} Normalized distributions of the 2D axial ratios $Q$ for intrinsically prolate (red dashed line) and oblate (black solid line) systems with $s=0.3$. Pure oblate and prolate systems have very distinctive 2D distributions at high $Q$, with the first approaching a plateau and the second dropping as steeply as $Q^{-3}$.}
\end{figure*}

\smallskip
These findings are in qualitative agreement with our predictions, but one should realize that the deprojection of shape is tricky, and the conclusions may be misleading if they are based on wrong assumptions concerning the shapes of the distributions. Using simple examples, we demonstrate here that the probability distribution of $Q$, $P(Q)$, cannot easily distinguish between a sample of oblate systems and a sample of prolate systems with a large variety of 3D axial rations. Only a sample of prolate systems with a narrow range of 3D axial ratios is clearly distinguishable in $P(Q)$.

\smallskip
We first use equation (\ref{eq:shape_2D} ) to derive ${\rm d}P/{\rm d}Q$ for samples of pure oblate systems and pure prolate systems with a constant $s=c/a$, and obtain
\begin{equation}\label{eq:2d_oblate}
\frac{{\rm d}P}{{\rm d}Q} = \frac{Q}{\sqrt{(1-s^2)(Q^2-s^2)}}\,\,\,,\qquad{\rm oblate}\,\,\,,
\end{equation}
\begin{equation}\label{eq:2d_prolate}
\frac{{\rm d}P}{{\rm d}Q }= \frac{s^2}{Q^2\,\sqrt{(1-s^2)(Q^2-s^2)}}\,\,\,,\qquad{\rm prolate}\,\,\,.
\end{equation}
The distribution is not defined at $Q<s$ in both cases. However, in the range $Q \gg s$, ${\rm d}P/{\rm d}Q$ flattens off into a constant for oblate systems, while it drops steeply as $\propto Q^{-3}$ for prolate systems.
This can be seen in Fig. \ref{fig:distr_pure_systems}, which shows ${\rm d}P/{\rm d}Q$ for pure oblate and prolate systems with $s=0.3$. Thus, samples of pure prolate or oblate systems with constant axial ratios yield very distinct 2D distributions of $Q$.

\begin{figure*}
\includegraphics[scale=0.4]{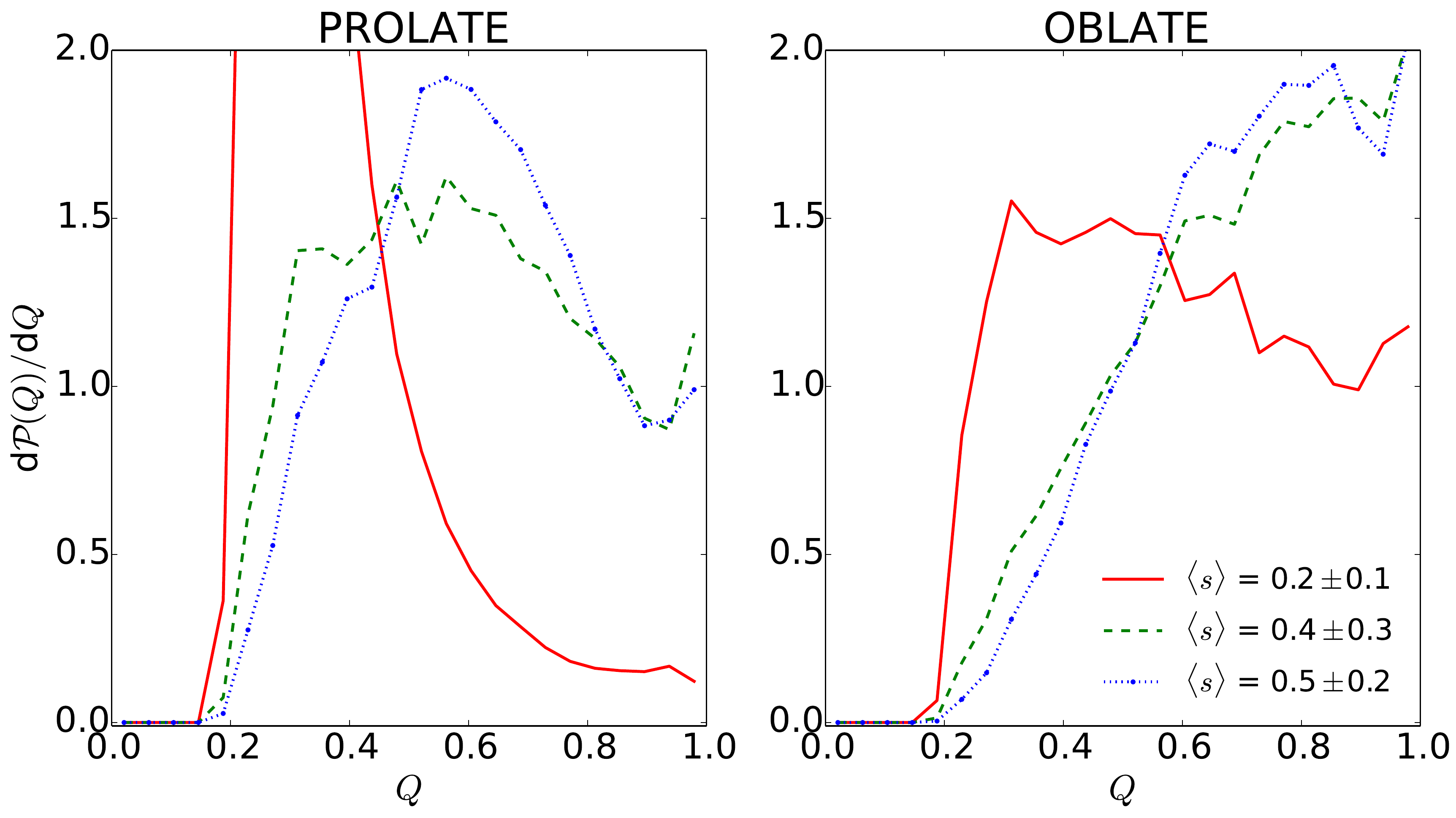}
\caption{\label{fig:prolate_oblate_q_distr} Distributions of the 2D axial ratios $Q$ for intrinsically prolate (left-hand panel) and oblate (right-hand panel) systems. The intrinsic axial ratios are distributed on a truncated Gaussian with mean and dispersion indicated in the legend. A very broad Gaussian distribution for a prolate system (left-hand panel, dot-dashed blue line) results in a ${\rm d}P/{\rm d}Q$ comparable to that of intrinsically oblate system (right-hand panel, solid red line).}
\end{figure*}

\smallskip
However, when considering a sample of galaxies with a broad distribution of intrinsic shapes, the resulting distribution of the projected axial ratios may not exhibit such strong features, and it may be more difficult to assess the underlying intrinsic axial ratio. For instance, Fig. \ref{fig:prolate_oblate_q_distr} shows ${\rm d}P/{\rm d}Q$ for prolate and oblate systems which have their $s$ randomly selected from a truncated Gaussian distribution (i.e. with $0.2\le s\le 1$) with mean and standard deviation as indicated in the legend. One can see that prolate systems with a large scatter have a resulting probability distribution of $Q$ that does not have a steeply declining part at high $Q$ and may therefore be confused with the distribution for oblate objects. For example, a sample of prolate systems with $s=0.4 \pm 0.3$ and a sample of oblate systems with $s=0.2 \pm 0.1$ show similar 2D distributions, with a slowly decreasing ${\rm d}P/{\rm d}Q$ at high $Q$.  On the other hand, the 2D distribution of a sample of oblate systems with either a large scatter (or a high $\langle s \rangle$) is gradually rising over a wide range of $Q$ values, with the flattening occurring only at very high $Q$. 
The features of ${\rm d}P/{\rm d}Q$ are even more problematic to distinguish when the distribution of the intrinsic shape parameters is not Gaussian, which is the case for the galaxies in our sample.

\begin{figure*}
\includegraphics[scale=0.35]{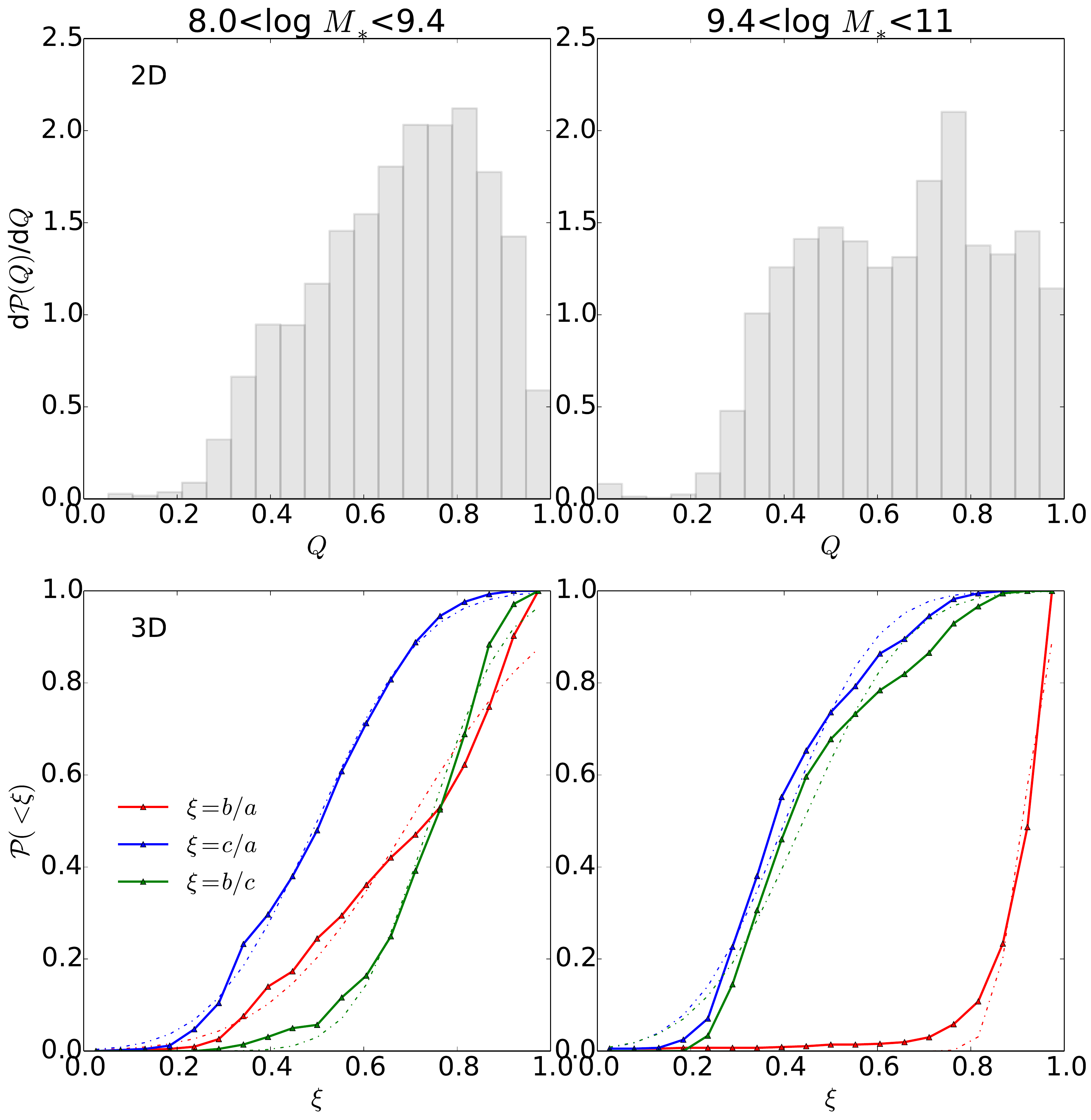}
\caption{\label{fig:observations_mass} Distributions of axial ratios in 2D (top) and 3D (bottom) in the simulations, in two mass bins. The dot-dashed lines in the bottom panel show a fit to a cumulative Gaussian distribution. The 2D distributions in the two mass bins are rather similar, despite the fact that the 3D distributions are very different.}
\end{figure*}

\smallskip
In order to obtain ${\rm d}P/{\rm d}Q$ for every snapshot in our sample of simulated galaxies we project the best-fitting intrinsic 3D ellipsoid on to 100 random lines of sight. Fig. \ref{fig:observations_mass} shows the results for two different mass bins and compares the 2D (top panels) and 3D (bottom panels) distributions of the axial ratios. One can see that about half the systems in the lowest mass bin are intrinsically triaxial ($q\sim p$), and the other half are prolate ($q<p$), but the distributions of the 3D axial ratios have large dispersions around the mean. Therefore, the resulting distribution of the 2D axial ratio is increasing over a wide range of $Q$ values, and it shows a steeply decreasing tail only at very high $Q$ values.  In the high-mass bin, the systems clearly tend to be oblate, with $q\sim 1$ and $p\ll1$, but the 2D distribution is rather similar to the one in the low-mass bin. If we fit the  distribution of the intrinsic parameters with a Gaussian distribution (dot-dashed lines), we find that in the low-mass bin $P(s)$ is well fitted with a Gaussian with mean $\sim0.5$ and standard deviation $\sim0.2$. This is the distribution of $s$ assumed for the blue line in Fig. \ref{fig:prolate_oblate_q_distr} (left-hand panel), which is qualitatively similar to the top-left panel in Fig. \ref{fig:observations_mass}.

\section{Conclusions}\label{sec:conclusions}
We have used cosmological simulations of galaxies at high redshift to study the evolution of the global shape of the components of DM, stars and gas. Our earlier analysis \citep{Tacchella+15b,Tacchella+15c,Zolotov+15} showed that every galaxy typically undergoes a major compaction event into a blue-nugget phase at $z\sim2-4$. This phase marks the transition from a DM-dominated central body to a self-gravitating baryonic core. The evolution of shape turns out to be strongly linked with this transition.

\noindent
Our results can be summarized as follows
\begin{description}

\item[$\bullet$]~At high redshift, in the DM-dominated phase, the stellar and DM systems are prolate-triaxial and mutually aligned. This confirms earlier results by \citet{Ceverino+15} but using a larger sample and performing a more thorough analysis.

\item[$\bullet$]~The early elongation is supported by an anisotropic velocity dispersion that results from the assembly of the galaxy via mergers and smoother streams along a dominant filament of the cosmic web.

\item[$\bullet$]~Torques exerted by the DM halo are capable of inducing the elongation of the stellar system and its alignment with the halo.

\item[$\bullet$]~In association with the transition to a self-gravitating compact core, small-pericentre orbits of DM and stars are scattered and the system evolves into a more spherical and oblate configuration, aligned with the gas disc and associated with rotation.

\item[$\bullet$]~While more massive galaxies (as ranked at a given redshift) typically compactify at a higher redshift, the transition typically occurs when the stellar mass is $\sim 10^{9.4}$ M$_\odot$ and the escape velocity from the core is $\sim 100$ km s$^{-1}$. This indicates that stellar feedback is effective at maintaining the DM dominance in the core.

\item[$\bullet$]~Feedback plays a role in the morphology of the galaxy. Simulations with no radiative feedback produce rounder galaxies because the transition to a baryon-dominated core happens at earlier time.
\item[$\bullet$]~The early elongated phase may drive compaction by generating angular-momentum loss, and the transition to the oblate phase may be instrumental in suppressing torque-driven inflows in the galaxy and thus contributing to the subsequent quenching in the core.
\end{description}
\smallskip
While the results deduced from our  cosmological simulations are robust,  some of the theoretical scenarios described above are tentative and yet to be properly worked out. This refers in particular to the way a prolate halo that dominates the central potential can induce an aligned prolate shape in the stellar system, and the way a central baryonic mass concentration can make the stellar system and the inner DM halo round.  
\smallskip
An observational confirmation of the 
predicted evolution of shape is not easy to obtain, and is beyond the scope of the current paper. We only comment the following. 
\begin{description}
\item[$\bullet$]~In particular, the observed distribution of projected axial ratios is capable of identifying a population of prolate systems with a narrow range of three-dimensional axial ratios, but cannot easily distinguish between oblate and prolate systems with a wide range of three-dimensional axial ratios.
\item[$\bullet$]~There is partial observational evidence for a prolate population of low-mass galaxies at high redshift, consistent with the simulation results \citep{Chang+13,vanderWel+14,Takeuchi+15}.
\end{description}
\smallskip
The transformation of global shape is thus one of the very interesting consequences of the wet compaction events that drive the evolution of galaxies in several different ways.  The compactions produce compact star-forming blue nuggets and may activate central AGNs. These trigger gas depletion from the centre and lead to inside-out quenching \citep{Tacchella+15b,Zolotov+15}. The sequence of compaction and quenching attempts can explain the confinement of star-forming galaxies into a narrow main sequence \citep{Tacchella+15c}.  The associated transformation of shape adds to the realization that the events of wet compaction are important milestones in the history of galaxies. 

\label{lastpage}

\section{Acknowledgement}

We acknowledge stimulating discussions with Marcella Carollo and David Koo. The simulations were performed at the National Energy Research Scientific Computing centre (NERSC), Lawrence Berkeley National Laboratory, and at NASA Advanced Supercomputing (NAS) at NASA Ames Research Center. The analysis has been performed on the Astric cluster at HU. This work was supported by  ISF grant 24/12, by BSF grant 2014-273,  by GIF grant G-1052-104.7/2009, by the I-CORE Program of the PBC,
ISF grant 1829/12, by CANDELS grant HST-GO-12060.12-A, and by NSF grants AST-1010033 and AST-1405962.

\bibliographystyle{mn2e}
\bibliography{biblio}

\end{document}